\documentclass[fleqn,usenatbib]{mnras}

\usepackage{amsmath}
\usepackage{amssymb}
\usepackage{tablefootnote}
\usepackage{newtxtext,newtxmath}

\usepackage[T1]{fontenc}

\DeclareRobustCommand{\VAN}[3]{#2}
\let\VANthebibliography\thebibliography
\def\thebibliography{\DeclareRobustCommand{\VAN}[3]{##3}\VANthebibliography}

\usepackage{graphicx}	
\usepackage{hyperref}
\usepackage{verbatim}
\usepackage{tabularx}
\usepackage{xspace}	
\usepackage{xcolor}

\newcommand{\unit}[1]{\ensuremath{\mathrm{\,#1}}\xspace}
\newcommand{\feh}{\unit{[Fe/H]}}
\newcommand{\teff}{\ensuremath{T_\mathrm{eff}}\xspace}
\newcommand{\logg}{\ensuremath{\log\,g}\xspace}
\newcommand{\alphafe}{\unit{[\alpha/Fe]}}
\newcommand{\gaia}{{\it Gaia}\xspace}
\newcommand{\kms}{\unit{km\,s^{-1}}}

\newcommand{\change}[1]{{{#1}}}

\title[DESI EDR MWS VAC]{DESI Early Data Release Milky Way Survey Value-Added Catalogue}

\author[S. E. Koposov et al.]{Sergey~E.~Koposov,$^{1,2,3}$\thanks{E-mail: skoposov@ed.ac.uk}
C.~Allende~Prieto,$^{4}$
A.~P.~Cooper,$^{5}$
T.~S.~Li,$^{6}$
L.~{Beraldo e Silva},$^{7,8}$
B.~Kim,$^{1}$
\newauthor
A.~Carrillo,$^{9}$
A.~Dey,$^{10}$
C.~J.~Manser,$^{11,12}$
F.~Nikakhtar,$^{13}$
A.~H.~Riley,$^{9}$
C.~Rockosi,$^{14,15,16}$
M.~Valluri,$^{7,17}$
\newauthor
J.~Aguilar,$^{18}$
S.~Ahlen,$^{19}$
S.~Bailey,$^{18}$
R.~Blum,$^{10}$
D.~Brooks,$^{20}$
T.~Claybaugh,$^{18}$
S.~Cole,$^{9}$
A.~de la Macorra,$^{21}$
\newauthor
B.~Dey,$^{22}$
J.~E.~Forero-Romero,$^{23,24}$
E.~Gaztañaga,$^{25,26,27}$
J.~Guy,$^{18}$
A.~Kremin,$^{18}$
L.~Le~Guillou,$^{28}$
\newauthor
Michael~E.~Levi,$^{18}$
M.~Manera,$^{29,30}$
A.~Meisner,$^{10}$
R.~Miquel,$^{31,30}$
J.~Moustakas,$^{32}$
J.~Nie,$^{33}$
\newauthor
N.~Palanque-Delabrouille,$^{34,18}$
W.~J.~Percival,$^{35,36,37}$
M.~Rezaie,$^{38}$
G.~Rossi,$^{39}$
E.~Sanchez,$^{40}$
E.~F.~Schlafly,$^{41}$
\newauthor
M.~Schubnell,$^{42,17}$
G.~Tarl\'{e},$^{17}$
B.~A.~Weaver,$^{10}$
and Z.~Zhou$^{33}$\\
Authors affiliations are provided in the Appendix \ref{sec:affiliations}
}

\date{Accepted XXX. Received YYY; in original form ZZZ}

\pubyear{2015}

\begin{document}
\label{firstpage}
\pagerange{\pageref{firstpage}--\pageref{lastpage}}
\maketitle

\begin{abstract}
We present the stellar value-added catalogue based on the Dark Energy Spectroscopic Instrument (DESI) Early Data Release. The catalogue contains radial velocity and stellar parameter measurements for $\simeq$ 400,000 unique stars observed during commissioning and survey validation by DESI.  These observations were made under conditions similar to the Milky Way Survey (MWS) currently carried out by DESI but also include multiple specially targeted fields, such as those containing well-studied dwarf galaxies and stellar streams. The majority of observed stars have $16<r<20$ with a median signal-to-noise ratio in the spectra of $\sim$ 20.
In the paper, we describe the structure of the catalogue, give an overview of different target classes observed, as well as provide recipes for selecting clean stellar samples. We validate the catalogue using external high-resolution measurements and show that radial velocities, surface gravities, and iron abundances determined by DESI are accurate to 1 km\,$s^{-1}$, $0.3$ dex and $\sim$ 0.15 dex respectively. We also demonstrate possible uses of the catalogue for chemo-dynamical studies of the Milky Way stellar halo and Draco dwarf spheroidal.  
The value-added catalogue described in this paper is the very first DESI MWS catalogue.  The next DESI data release, expected in less than a year, will add the data from the first year of DESI survey operations and will contain approximately 4 million stars, along with significant processing improvements.
\end{abstract}



\section{Introduction}
\label{sec:intro}

In the last twenty years, major progress in stellar astrophysics, Galactic dynamics and Galactic archaeology has been driven to a large degree by the arrival of massive digital surveys. The first surveys were mainly photometric, such as 2MASS \citep{skrurskie2006}, SDSS-I \citep{york2000} and PanSTARRS \citep{chambers2016}. More recently, these datasets have been complemented by unprecedented astrometric measurements from {\it Gaia}, which provided high-quality proper motions and parallaxes \citep{gaia2021}. Over the last ten years or so, an increasingly large number of surveys such as SEGUE \citep{yanny2009, rockosi2022}, LAMOST \citep{zhao2012}, GALAH \citep{desilva2015}, APOGEE \citep{majewski2017}, Gaia-ESO \citep{gilmore2022} and {\it Gaia} itself \citep{katz2023} have provided spectroscopic data for hundreds of thousands, and in some cases millions, of stars. These surveys have made particularly valuable contributions to our understanding of the Galaxy and its constituent stellar populations. The combination of radial velocities with proper motions provides full 6-D coordinates that enable, for example, the discovery of ancient substructure in "integral of motion" spaces \citep{myeong2018,koppelman2019,naidu2020} and measurements of the distribution of mass in the Milky Way \citep{deason2021,bird2022,koposov2023}. The measurement and interpretation of chemical abundances on the scale of these surveys is a rapidly developing field that has already provided significant insight into the formation and assembly of the Milky Way \citep{horta2021,hayden2015,mackareth2019,belokurov2022}. Finally, the large numbers of stars in these surveys allow for many serendipitous discoveries, including extremely metal-poor stars \citep{caffau2011,allendeetal23-1} and hyper-velocity stars \citep{koposov2020}. These successes have motivated a new generation of even more ambitious stellar spectroscopic surveys such as DESI, 4MOST, SDSS-V and WEAVE \citep{dejong2019,Cooper23,kollmeier2017,jin2023}.

In this contribution, we describe the first catalogue of measurements from the stellar part of the Dark Energy Spectroscopic Instrument (DESI) Survey's Early Data Release \citep[EDR;][]{desi_edr}. DESI is a 5020-fiber, $3.2^\circ$-diameter field-of-view, moderate-resolution spectrograph ($2000<\lambda/\Delta\lambda<5000$ for 3600\AA\ $<\lambda<$ 9800\AA~spread across three arms) mounted at the prime focus of the Mayall 4\,m telescope at the Kitt Peak National Observatory \citep{desi2016a,DESI_Instr_Overview2022}. 
 While DESI's primary mission is to provide sub-percent
cosmological measurements of the expansion history of the universe, 
the instrument is also
being used for a Milky Way Survey \citep[DESI-MWS;][]{Cooper23}
of $\approx7$~million stars with $r\lesssim 19$~mag, focused on the
thick disc and stellar halo of our Galaxy. DESI-MWS will also create the
largest spectroscopic catalogue to date of white dwarfs, stars within 100~pc of the Sun, RR Lyrae and BHB stars up to distances of 100 kpc. Serendipitous discoveries have already been made using DESI-MWS data, with the identification of metal-poor stars and rare magnetic white dwarfs exhibiting emission features \citep{allendeetal23-1, manseretal23-1}


Following the first light with the instrument in October 2020 and the initial period of commissioning, 
DESI  moved into a
Survey Validation (SV) phase \citep{desi_sv}.
During this period (approximately Dec 2020 to June 2021), the DESI Collaboration observed specific areas of sky (referred to as tiles) with a variety of instrument configurations and target selection schemes to validate observing and data reduction procedures and to optimise the definition of the primary survey target classes (i.e.,
Milky Way Stars and white dwarfs, \citealt{allendeetal20-1,Cooper23}; Bright Galaxies, \citealt{ruiz-maciasetal20-1,hahn22a}; Luminous Red Galaxies, \citealt{zhouetal20-1,zhouetal23-1}; Emission-Line
Galaxies, \citealt{raichooretal20-1, raichooretal23-1}; and QSOs, \citealt{yecheetal20-1, chaussidonetal23-1}). The final definitions of all the DESI target classes are summarised in \citet{myers22a}. The SV phase was split into 3
campaigns: the first two took an iterative approach
to optimise the target selection and survey strategy. During the third campaign,
DESI conducted a ``One-Percent Survey'', covering approximately 140~deg$^2$
of sky (i.e., 1 per cent of the overall DESI footprint) using the same target selection as the main DESI survey but with much higher completeness across different target classes. After
 completing the SV phase, DESI began its main 5-year survey program in May 2021.

In this paper, we describe the observation and analysis of  
stellar spectra taken during the DESI commissioning and SV campaigns, supplementing and updating the description given in \citet{Cooper23}, and provide a guide to the Milky Way Survey Value-Added Catalogue\footnote{\url{https://data.desi.lbl.gov/doc/releases/edr/vac/mws/}} (MWS VAC) constructed from stellar spectra in the DESI EDR.
In Section~\ref{sec:dataset}, we give an overview of the target selections used in each campaign.  Section~\ref{sec:processing} describes the MWS-specific data analysis
pipelines that measure stellar radial velocities and atmospheric parameters.  Section~\ref{sec:dataproducts} presents the individual data products provided in the MWS VAC. In Section~\ref{sec:combined_catalogs}, we describe the combined catalogue of measurements, which we expect to be the most convenient data product for general use.
Section~\ref{sec:analysis} presents several use cases that
exemplify scientific applications of the EDR data, and Section~\ref{sec:limitations} describes known limitations of the MWS VAC.

\section{Dataset} 
\label{sec:dataset}

The DESI EDR comprises observations obtained during commissioning and survey validation, which began in December 2020 and ended in June 2021. As described in \citet{desi_edr}, the data are organised by survey and program. Table~\ref{tab:survey_program} provides the list of survey and programme combinations with the number of stars in the MWS VAC for each combination.

The surveys, CMX, SV1, SV2 and SV3, correspond to commissioning (CMX) and several stages of survey validation (SV1, SV2 and SV3), respectively. The SV3 data are the closest in target selection to those in the main DESI survey.

The programs, such as `dark', `bright' and `backup', correspond to the observational conditions. During a given survey, DESI  conducts observations in different programs based on a metric referred to as survey speed \citep{schlafly2023}. The SV1 program called `other' is a special category comprising observations dedicated to specific validation tasks. The main MWS survey is carried out exclusively in the `bright' program. However, during survey validation, a large number of targets in specific fields and special classes of stellar targets outside the main MWS selection function were observed in `dark' and `other' programs. 

The target selection evolved as DESI progressed from commissioning and survey validation up to the start of the main survey. Additionally, new target classes and customised target selection criteria were introduced for specific fields during the SV programs. In the following sections, we briefly summarise different target classes relevant to Milky Way science.

\begin{table}
\caption{Different DESI survey and program combinations that are in the EDR data release, together with the number of objects included in the MWS VAC.}
\begin{center}
    
 \begin{tabular}{llr}
 {Survey} & {Program} & 
 {Number of stars} \\
\hline
CMX & other & 728 \\
SPECIAL & dark & 2457 \\
SV1 & backup & 60,959\\
SV1 & bright & 45,738\\
SV1 & dark & 41,984 \\
SV1 & other & 82,247 \\
SV2 & backup & 2953\\
SV2 & bright & 7217\\
SV2 & dark & 3956\\
SV3 & backup & 83,976 \\
SV3 & bright & 236,448 \\
SV3 & dark & 57,025 \\
\hline
Total &    & 625,688 \\
\hline
 \end{tabular}
\end{center}
\label{tab:survey_program}
\end{table}

\subsection{Targets}

Details of the DESI target selection and different target classes are given in \citet{myers22a, desi_sv, desi_edr, Cooper23}. Here, we provide a summary relevant to Milky Way science and necessary for the interpretation of the value-added catalogue.

During commissioning, targeting information was not carefully tracked. Since the total number of spectra from commissioning included in the DESI EDR is very small, we do not discuss them here. After commissioning, each of the SV1, SV2 and SV3 surveys used distinct target selection strategies in which new target classes were introduced, and some selections were modified or renamed.

The majority of stellar spectra in the EDR were observed as MWS targets, flux calibration standards, or one of several classes of `secondary' targets. Secondary program targets include those in several specific fields (for example, the M31 observations described by \citealt{Dey2023}) and spare fibre programs used in multiple SV fields.

In the MWS VAC, as in other DESI data products, information about why a given source was targeted is encoded in the
integer columns (bitmasks) {\tt DESI\_TARGET}, {\tt MWS\_TARGET} and {\tt SCND\_TARGET}.
 The {\tt DESI\_TARGET} bitmask provides high-level separation between the different components of the DESI survey, distinguishing between targets associated with the primary extra-galactic and stellar programs, the secondary programs, standard stars, and sky fibres; {\tt MWS\_TARGET} is a detailed bitmask for the Milky Way Survey, separating different classes of Milky Way target;  likewise, {\tt SCND\_TARGET} bitmask separates different classes of secondary targets (both stellar and extra-galactic). In the SV programs, many secondary classes correspond to subsets of the main MWS survey target selection that were given higher priority to increase the number of representative spectra for particular types of stars.

Since the SV1, SV2 and SV3 surveys used different target selections, the MWS VAC has three distinctly named target bitmasks for each survey; for example,  {\tt SV1\_DESI\_TARGET},  {\tt SV1\_MWS\_TARGET} and  {\tt SV1\_SCND\_TARGET} are the bitmasks for the SV1 survey.  Table~\ref{tab:sv_masks} specifically lists the bitmask names for SV1, SV2 and SV3. Stars observed in a given SV program should be selected using the corresponding set of target bitmasks.

\begin{table}
\begin{center}
 \begin{tabular}{ll}
 \hline
 Targeting column names\\
 \hline
 {SV1\_DESI\_TARGET}\\
 {SV1\_MWS\_TARGET}\\
 {SV1\_SCND\_TARGET}\\
 \hline
 {SV2\_DESI\_TARGET}\\
 {SV2\_MWS\_TARGET}\\
 {SV2\_SCND\_TARGET}\\
 \hline
 {SV3\_DESI\_TARGET}\\
 {SV3\_MWS\_TARGET}\\
 {SV3\_SCND\_TARGET}\\
 \hline
\end{tabular}
\end{center}
\caption{The name of the columns that contain selection bit masks for SV1, SV2 and SV3 surveys.}
\label{tab:sv_masks}
\end{table}
\subsection{High-level DESI target types}

The {\tt DESI\_TARGET} bitmask encodes the overall DESI survey component for which each source was observed and also separates science targets from flux calibration standard stars and sky fibres. Table~\ref{tab:main_desi_target} lists the subset of {\tt DESI\_TARGET} bits corresponding to standard stars, MWS targets and secondary targets. These bits have the same meaning for SV1, SV2 and SV3, although the actual selections did change somewhat\footnote{The magnitude range for {\tt STD\_BRIGHT} was changed from $15 \le G < 18$ in SV1 to $16 \le G < 18$ in SV2 and SV3. The colour criteria and the magnitude range of {\tt STD\_FAINT} were the same in SV1, SV2 and SV3.}. The {\tt DESI\_TARGET} bitmask for a given SV stage is nonzero only for stars observed in that stage. For example, only stars observed in SV1 have a non-zero value of {\tt SV1\_DESI\_TARGET}\footnote{The VAC also has a small number (759) of rows with zero {\tt SV1\_DESI\_TARGET}, {\tt SV2\_DESI\_TARGET} and {\tt SV3\_DESI\_TARGET}, but non-zero {\tt DESI\_TARGET}. These are fibres targeted as sky but having significant flux in them.}.

\begin{table}
\caption{{\tt DESI\_TARGET} bits for target classes relevant for MWS (the bits are identical for SV1, SV2 and SV3)} \begin{tabular}{ccc}
\hline
{Bit name} & 
{Bit index} &
{Description}
\\
\hline
{\tt STD\_FAINT}            & 33 &   Standard stars for dark/grey conditions \\
{\tt STD\_WD}               & 34 &   White Dwarf standard stars \\
{\tt STD\_BRIGHT}           & 35 &   Standard stars for bright conditions \\
{\tt MWS\_ANY}       & 61 & Any MWS bit is set \\
{\tt SCND\_ANY}      & 62 & Any secondary bit is set \\
\hline
\end{tabular}
\label{tab:main_desi_target}
\end{table}

\begin{table*}
\caption{Milky Way Survey targeting bits for the SV1.}
\begin{tabular}{ccc}
\hline
{Bit name} & 
{Bit index} &
{Description} 
\\
\hline
{\tt MWS\_MAIN\_BROAD}        & 0  & MWS magnitude-limited bulk sample \\ 
{\tt MWS\_WD}                 & 1  & MWS White Dwarf target            \\
{\tt MWS\_NEARBY}             & 2  & MWS volume-complete ($\sim$100\,pc) sample \\
{\tt MWS\_MAIN\_BROAD\_NORTH} & 4  & MWS targets from Bok/Mosaic       \\
{\tt MWS\_MAIN\_BROAD\_SOUTH} & 5  & MWS targets from DECam            \\
{\tt MWS\_BHB}                & 6  & MWS Blue Horizontal Branch targets                          \\
{\tt MWS\_MAIN\_FAINT}        & 14 & MWS magnitude-limited sample                   \\
{\tt MWS\_MAIN\_FAINT\_NORTH} & 15 & MWS magnitude-limited sample from Bok/Mosaic   \\
{\tt MWS\_MAIN\_FAINT\_SOUTH} & 16 & MWS magnitude-limited sample from DECam        \\
{\tt GAIA\_STD\_FAINT}        & 33 & {\it Gaia}-based standard stars for dark/grey conditions   \\
{\tt GAIA\_STD\_WD}           & 34 & {\it Gaia}-based White Dwarf stars for use as standards              \\
{\tt GAIA\_STD\_BRIGHT}       & 35 & {\it Gaia}-based standard stars for bright conditions                \\
{\tt BACKUP\_GIANT\_LOP} &    58  & {\it Gaia}-based possible giant stars  for backup survey \\
{\tt BACKUP\_GIANT} &        59  & {\it Gaia}-based selection of giant stars for backup survey   \\
{\tt BACKUP\_BRIGHT}          & 60 & Bright backup {\it Gaia} targets                     \\
{\tt BACKUP\_FAINT}           & 61 & Fainter backup {\it Gaia} targets                    \\
{\tt BACKUP\_VERY\_FAINT}     & 62 & Even fainter backup {\it Gaia} targets               \\
\hline
\end{tabular}

\label{tab:sv1_mws_targets}
\end{table*}

\subsection{MWS survey targets}

The criteria used to select stars for the Milky Way Survey (i.e.\ those with the {\tt MWS\_ANY} bit set in the {\tt DESI\_TARGET} bitmask) were different in SV1 and SV2. The SV3 observations used the same target classes and selections as SV2. The MWS target classes defined for SV1 are given in Table~\ref{tab:sv1_mws_targets}, while the new or modified SV2 target classes are given in Table~\ref{tab:sv2_mws_targets}. The exact criteria that define these classes in the main DESI survey are provided in \citet{Cooper23,myers22a}. Note that the criteria used to define a given class in SV may not be the same as those used for the main survey class with the same name, although in most cases, they are very similar. \change{To make it easier to interpret the catalogues we also summarise the selection criteria for the main (most numerous) MWS classes for the SV3 data in the appendix.} Briefly, these classes are as follows. 

\begin{itemize}
    \item {\tt MWS\_MAIN\_BROAD} class correspond to a magnitude limited selection in the range $16<r<19$ without kinematic or colour cuts, supplemented by basic photometric quality requirements and was used in SV1 only \footnote{Several target classes (e.g.\ {\tt MWS\_BROAD}) have {\tt NORTH} and {\tt SOUTH} sub-categories (e.g.\ {\tt MWS\_BROAD\_NORTH} and {\tt MWS\_BROAD\_SOUTH}). Sources selected using Bock/Mosaic photometry in the northern section of the DESI survey footprint have the {\tt NORTH} bit set, while those selected using DECam photometry in the southern section of the footprint have the {\tt SOUTH} bit set. Since the MWS selection criteria were identical for the northern and southern parts of the footprint, these subclasses only indicate the origin of the input photometry.}. 
    \item {\tt MWS\_MAIN\_FAINT} corresponds to faint stellar targets with $19<r<20$ observed during SV1 only. \change{We note however that in the main survey this target category is observed as a secondary target, rather than the main survey target.}
    \item {\tt MWS\_MAIN\_BLUE} comprises stars with $g-r<0.7$. This class has no proper motion or parallax selection and was used starting from SV2. 
    \item {\tt MWS\_MAIN\_RED} comprises stars selected using proper motion and parallax with $g-r>0.7$, intended to favour the selection of distant giants. This category was used starting from SV2.
    \item {\tt MWS\_BROAD} class correspond to a magnitude limited selection in the range $16<r<19$ with $g-r>0.7$, but not satisfying astrometric cuts of {\tt MWS\_MAIN\_RED} and was used starting from SV2.
     \item {\tt MWS\_NEARBY} selects stars in a volume-limited sample within 100\,pc, using \gaia parallaxes, and was used starting from SV1.
    \item {\tt MWS\_WD} corresponds to white dwarfs and was used starting from SV1 \footnote{Different selections were used in different SVs and will be described in the dedicated paper}
    \item {\tt MWS\_BHB} comprises blue horizontal branch stars selected using the DESI Legacy Survey photometry and was used from SV1. We note that this category, which is described in \citep{Cooper23}, is distinct from the {\tt BHB} secondary target category defined for observations of more distant BHB candidates in the dark time (see below).
    \item {\tt GAIA\_STD\_WD}, {\tt GAIA\_STD\_FAINT}, {\tt GAIA\_STD\_BRIGHT} are flux calibration standards selected based on \gaia data alone. These are used in areas without the DECaLS photometry to select regular DESI standards. They were included starting from SV1.
    \item {\tt BACKUP\_BRIGHT}, {\tt BACKUP\_FAINT}, {\tt BACKUP\_VERY\_FAINT}, {\tt BACKUP\_GIANT}, {\tt BACKUP\_GIANT\_LOP} are targets that were observed when conditions were too poor for the bright or dark SV programs. The exact definition of the backup survey and its target classes will be described in a follow-up paper. Backup targets were included starting from SV1.
 \end{itemize}

\change{For all the target classes listed above in Tables ~\ref{tab:sv1_numbers}, \ref{tab:sv2_numbers}, \ref{tab:sv3_numbers} we provide the number of objects in these classes  observed in different surveys and programmes  included in this VAC. Note that the same object may have more than one target class and may have been observed in multiple surveys.}

\change{In this paper we do not describe the selection function of the survey in detail, partially because this VAC is based on the early data release which is a combination of commissioning and science verification surveys and the target selection was changing throughout. However, for each survey and programme the information needed to reproduce the selection function or replay the target selection is available as part of the DESI EDR. Specifically:
\begin{itemize}
    \item The details of the  target selection for different classes are given in \citet{myers22a}
    \item The code used for target selection in SV1, SV2, SV3 and main survey is publicly available \url{https://desitarget.readthedocs.io/en/latest/} and is based on public Legacy Survey photometry data.
    \item The list of all potential survey targets selected by the DESI targeting code  is available as part of the DESI EDR. I.e. the list of all the potential targets of different classes for SV3 is available here \url{https://data.desi.lbl.gov/public/edr/target/catalogs/dr9/0.57.0/targets/sv3/resolve/}. These are all the objects that DESI considered targeting, before the fiber assignment on individual DESI tiles was run.
\end{itemize}

To give an idea of how this can be used, Table~\ref{tab:selection_frac} provides the fraction of objects of different MWS target types in the SV3 footprint that ended up being  observed in SV3 in  bright programme and are in this VAC.  These numbers were obtained by taking the ratio of  number of objects in different target categories in the VAC divided by the number of objects in the same target class in the SV3 footprint in the input catalogues. The reason why those fractions are not 100\% is primarily because of the fiber allocation, as the number of potential targets can be larger than the number of fibers and because there is competition between fiber allocation to sky, standard stars, BGS program targets and MWS targets. The table shows that for the main MWS classes in SV3 survey  the completeness of observations was above 70\%. 


}

\begin{table*}
\caption{Additional and updated names and targeting bits for the Milky Way Survey used during SV2.}
\begin{tabular}{ccc}
\hline
{Bit name} & 
{Bit index} &
{Description}
\\
\hline
{\tt MWS\_BROAD}             & 0 & MWS magnitude-limited bulk sample      \\
{\tt MWS\_BROAD\_NORTH}      & 4 & MWS cuts in Bok/Mosaic imaging footprint  \\
{\tt MWS\_BROAD\_SOUTH}      & 5 & MWS cuts in DECam imaging footprint     \\
{\tt MWS\_MAIN\_BLUE}        & 8 & MWS magnitude-limited blue sample \\
{\tt MWS\_MAIN\_BLUE\_NORTH} & 9 & MWS magnitude-limited blue sample in Bok/Mosaic imaging footprint\\
{\tt MWS\_MAIN\_BLUE\_SOUTH} & 10 & MWS magnitude-limited blue sample in DECam imaging footprint \\
{\tt MWS\_MAIN\_RED}         & 11 & MWS magnitude-limited red sample \\
{\tt MWS\_MAIN\_RED\_NORTH}  & 12 & MWS magnitude-limited red sample in Bok/Mosaic imaging footprint \\
{\tt MWS\_MAIN\_RED\_SOUTH}  & 13 & MWS magnitude-limited red sample in DECam imaging footprint\\
\hline
\end{tabular}
\label{tab:sv2_mws_targets}
\end{table*}

\begin{table}
\caption{The number of objects in different MWS target categories observed in SV1 and included in the VAC.}
\begin{center}
    
\begin{tabular}{cc}
\hline 
Target class & Number \\
\hline 
\multicolumn{2}{l}{Program: dark}  \\ \hline
MWS\_MAIN\_BROAD &  15638 \\
MWS\_WD &  652 \\
MWS\_NEARBY &  71 \\
MWS\_BHB &  22 \\
MWS\_MAIN\_FAINT &  3461 \\
GAIA\_STD\_FAINT &  4 \\
GAIA\_STD\_WD &  6 \\
BACKUP\_BRIGHT &  1 \\
BACKUP\_FAINT &  18 \\
BACKUP\_VERY\_FAINT &  14 \\
\hline 
\multicolumn{2}{l}{Program: bright}  \\ \hline
MWS\_MAIN\_BROAD &  32507 \\
MWS\_WD &  24 \\
MWS\_NEARBY &  112 \\
MWS\_BHB &  3 \\
MWS\_MAIN\_FAINT &  4893 \\
GAIA\_STD\_FAINT &  24 \\
GAIA\_STD\_BRIGHT &  170 \\
BACKUP\_BRIGHT &  490 \\
BACKUP\_FAINT &  123 \\
BACKUP\_VERY\_FAINT &  8 \\
\hline \multicolumn{2}{l}{Program: other}  \\ \hline
MWS\_MAIN\_BROAD &  44504 \\
MWS\_WD &  272 \\
MWS\_NEARBY &  158 \\
MWS\_BHB &  116 \\
MWS\_MAIN\_FAINT &  8028 \\
GAIA\_STD\_FAINT &  657 \\
GAIA\_STD\_WD &  22 \\
GAIA\_STD\_BRIGHT &  1540 \\
BACKUP\_BRIGHT &  1045 \\
BACKUP\_FAINT &  2868 \\
BACKUP\_VERY\_FAINT &  159 \\
\hline \multicolumn{2}{l}{Program: backup}  \\ \hline
MWS\_MAIN\_BROAD &  59456 \\
MWS\_WD &  494 \\
MWS\_NEARBY &  180 \\
MWS\_BHB &  48 \\
MWS\_MAIN\_FAINT &  289 \\
GAIA\_STD\_FAINT &  133 \\
GAIA\_STD\_WD &  34 \\
GAIA\_STD\_BRIGHT &  131 \\
BACKUP\_FAINT &  1136 \\
\end{tabular}
\end{center}
\label{tab:sv1_numbers}
\end{table}

\begin{table}
\caption{The number of objects in different MWS target categories observed in SV2 and included in the VAC.}
\begin{center}
    
\begin{tabular}{cc}
\hline \multicolumn{2}{l}{Program: dark}  \\ \hline
MWS\_BROAD &  2 \\
MWS\_WD &  270 \\
MWS\_NEARBY &  30 \\
MWS\_MAIN\_BLUE &  3315 \\
MWS\_MAIN\_RED &  12 \\
\hline \multicolumn{2}{l}{Program: bright}  \\ \hline
MWS\_BROAD &  1381 \\
MWS\_WD &  255 \\
MWS\_NEARBY &  53 \\
MWS\_BHB &  7 \\
MWS\_MAIN\_BLUE &  4727 \\
MWS\_MAIN\_RED &  724 \\
\hline \multicolumn{2}{l}{Program: backup}  \\ \hline
GAIA\_STD\_FAINT &  329 \\
GAIA\_STD\_BRIGHT &  217 \\
BACKUP\_BRIGHT &  1261 \\
BACKUP\_FAINT &  1114 \\
BACKUP\_VERY\_FAINT &  578 \\
\end{tabular}
\end{center}
\label{tab:sv2_numbers}
\end{table}

\begin{table}

\caption{The number of objects in different MWS target categories observed in SV3 and included in the VAC.}
\begin{center}
\begin{tabular}{cc}
\hline 
Target class & Number \\
\hline \multicolumn{2}{l}{Program: dark}  \\ \hline
MWS\_BROAD &  10 \\
MWS\_WD &  662 \\
MWS\_NEARBY &  40 \\
MWS\_BHB &  321 \\
MWS\_MAIN\_BLUE &  10928 \\
MWS\_MAIN\_RED &  26 \\
\hline \multicolumn{2}{l}{Program: bright}  \\ \hline
MWS\_BROAD &  87772 \\
MWS\_WD &  840 \\
MWS\_NEARBY &  355 \\
MWS\_BHB &  439 \\
MWS\_MAIN\_BLUE &  99448 \\
MWS\_MAIN\_RED &  25330 \\
\hline \multicolumn{2}{l}{Program: backup}  \\ \hline
GAIA\_STD\_FAINT &  19476 \\
GAIA\_STD\_BRIGHT &  14633 \\
BACKUP\_FAINT &  56663 \\
BACKUP\_VERY\_FAINT &  27185 \\
\end{tabular}
\end{center}

\label{tab:sv3_numbers}
\end{table}

\begin{table}
\caption{Fraction of the targets in the input catalogue within  the SV3/bright footprint that were observed in SV3 in bright program and have a measurement in the EDR VAC.}
\begin{center}    
\begin{tabular}{cc}
\hline
Target class & Fraction \\
\hline
     MWS\_BROAD &  0.77 \\
MWS\_WD &  0.95 \\
MWS\_NEARBY &  0.87 \\
MWS\_BHB &  0.84 \\
MWS\_MAIN\_BLUE &  0.83 \\
MWS\_MAIN\_RED &  0.83 \\
\hline
\end{tabular}
\end{center}
\label{tab:selection_frac}
\end{table}

\subsection{Secondary targets}
\label{sec:secondary_targets}

Alongside the main survey target classes, a variety of sources were observed as secondary targets. Some of these were defined to fill spare fibres, while others correspond to targets of particular interest on specific tiles observed during SV. 

Table~\ref{tab:sv_secondary_targets} lists the different secondary stellar target categories and the corresponding {\tt SCND\_TARGET} bit values. The table also gives the number of objects of each type included in the VAC. \change{Note that many of these object categories may have been observed as regular survey targets.} We briefly summarise these secondary classes here; more detailed descriptions are given in  \citet{myers22a,desi_sv,Cooper23,Dey2023}.

\begin{itemize}
\item {\tt WD\_BINARIES\_DARK}, and {\tt WD\_BINARIES\_BRIGHT} select white dwarf binary candidates using {\it Gaia} and GALEX photometry to assemble an unbiased, homogeneous sample for studying the evolution of close white dwarf binaries.
\item {\tt M31\_QSO}, {\tt M31\_STAR} and {\tt M31\_KNOWN} are targets selected for the dedicated M31 observations described in detail by \citet{Dey2023}.
\item  {\tt MWS\_DDOGIANTS} was originally intended for selecting highly probable halo giants using DDO photometry; however, this bit was not used for any DESI observations.
\item {\tt BHB} targets were selected as faint candidate horizontal branch stars, potentially up to distances of 150 kpc from the Galactic Centre. This class has a fainter magnitude limit than the regular survey {\tt MWS\_BHB} targets (see above) and was observed in dark conditions only. The targets were selected using Legacy survey $g-r$,$r-z$ colours selection analogous to Eqn. 6 in \citet{li2019} and include faint RR Lyrae from {\it Gaia} DR2 \citep{clementini2019} and \citet{sesar2017}. 
\item  {\tt FAINT\_HPM}, {\tt BRIGHT\_HPM} and {\tt HPM\_SOUM} targets were selected to identify fast-moving stars in the Milky Way based on parallax and proper motion. The goal of this selection was to identify rare populations of hypervelocity stars and SNIa remnants.
\item {\tt MWS\_CLUS\_GAL\_DEEP} is a target bit for a secondary program observing Milky Way star clusters and dwarf galaxies. During SV, dedicated observations were taken of the Draco and Ursa Major II dwarf spheroidal galaxies,  the open cluster M44 (Beehive/Praesepe) and the globular cluster pair M53 and NGC 5053.
Different combinations of colour and magnitude from Legacy Survey DR9 and parallax and proper motion from \gaia EDR3 were used in each case to select candidate members of these objects. These targets were only included in dedicated tiles observed in dark time and were given higher priority than all other MWS targets on those tiles, except white dwarfs. Note that additional tiles in the same areas were also observed under regular bright-time conditions (see {\tt MWS\_MAIN\_CLUSTER\_SV} below). 

\item {\tt MWS\_MAIN\_CLUSTER\_SV} has a similar purpose to the {\tt MWS\_MAIN\_CLUS\_GAL\_DEEP} class. It was used in a series of dedicated fields observed in SV1 to prioritise likely member stars of clusters and dwarf galaxies. 
These targets are, on average, brighter than those selected in {\tt MWS\_MAIN\_CLUS\_GAL\_DEEP}, and the selection was based on \gaia DR2. They were observed under both bright and dark conditions. In addition to the fields listed for {\tt MWS\_MAIN\_CLUS\_GAL\_DEEP}, this class was used to target stars in the globular clusters M5, M13, M92 and NGC 2419. A handful of these targets also appear in the outskirts of one SV3 field near M13.

\item {\tt MWS\_RRLYR}\footnote{Note that the main survey uses a different bit and different name for the RR Lyrae selection (bit 37 of the secondary targeting mask with the name {\tt MWS\_RR\_LYRAE}) } are RR Lyrae variables that were selected from the {\it Gaia} DR2 RRL  catalogue \citep{clementini2019}. Note that there is some overlap with the RRL targeted with the {\tt BHB} class. In the EDR, 42 stars out of 332 {\tt MWS\_RRLYR} targets are also {\tt BHB} targets.

\item {\tt MWS\_CALIB} and {\tt BACKUP\_CALIB} are classes used during SV only to prioritise stars with data from one or more previous surveys, including APOGEE \citep{apogee_dr17}, BOSS \citep{boss2013}, GALAH \citep{desilva2015} and Gaia-ESO 
\citep{gilmore2022}. The {\tt MWS\_CALIB} stars have \gaia $G$-band magnitudes $16<G<19$  while the {\tt BACKUP\_CALIB} stars have $G<16$.
\end{itemize}

\begin{table}
\caption{Secondary targeting bits relevant to MWS and the number of objects with those bits in the catalogue. Detailed descriptions of these target classes are given in the text.}
\begin{center}
\begin{tabular}{ccc}
\hline
Bit-name &  Bit index  & Number of objects\\
\hline
{\tt M31\_KNOWN} &            6  & 25   \\
{\tt M31\_QSO} &              7  & 15 \\ 
{\tt M31\_STAR} &             8  & 827 \\
{\tt MWS\_DDOGIANTS} &                         9 &   0 \\
{\tt MWS\_CLUS\_GAL\_DEEP} & 10  & 4405 \\ 
{\tt FAINT\_HPM} &                            12 & 371   \\
{\tt BHB} &                                   18  & 249 \\
{\tt HPM\_SOUM} &                      27 & 31  \\
{\tt MWS\_CALIB} &                     36 & 2515 \\
{\tt BACKUP\_CALIB} &                  37  & 2540 \\
{\tt MWS\_MAIN\_CLUSTER\_SV} &         38 & 4669 \\
{\tt MWS\_RRLYR} &                     39 & 332 \\
{\tt BRIGHT\_HPM} &                                40  & 5 \\
{\tt WD\_BINARIES\_BRIGHT} &                       41  & 71 \\
{\tt WD\_BINARIES\_DARK} &                         42  & 88 \\
\hline
\end{tabular}
\end{center}
\label{tab:sv_secondary_targets}
\end{table}

\subsection{Dedicated fields}
\label{sec:dedicated_fields}
The DESI SV program observations used either a generic target selection (preliminary versions of the target selection for the main survey) or field-specific selections when observing various objects of particular value to validate the DESI data reduction pipelines. 

In Table~\ref{tab:mws_sv1_areas}, we provide a list of the SV fields that are of most interest for the MWS. These include the fields designed for tests of the MWS strategy and pipelines listed in \citet{Cooper23} and several other fields of interest, such as a field centred on  M31 \citep{Dey2023} and additional observations of streams, clusters and dwarf galaxies. As mentioned above, the target selection in these fields was customised by adding additional secondary target classes with high priority for fibre assignment (see Section~\ref{sec:secondary_targets}). Most of these dedicated fields were observed in SV1, under the program `other'. Many of these fields are centred on stellar streams and dwarf galaxies and contain tens to hundreds of stars associated with those objects. These data will be described in upcoming publications. The data obtained from SV fields covering the Draco dwarf spheroidal are demonstrated in Section~\ref{sec:draco}.

\begin{table}
\caption{Dedicated observations in SV1 relevant to the Milky Way Survey. 
From left to right, columns give the name of the object, the central coordinates of the field, the total exposure time and the DESI tile number. 
}
\begin{tabular}{lllll}
\hline
Object &
{R.A. [$^\circ$] }  & 
{Decl. [$^\circ$]} & 
{$T_\mathrm{exp} (s)$ }& 
{Tile number}
\\
\hline
NGC 2419 & 114.221 & 38.469 & 9779.6 & 80618 \\
NGC 2419 & 114.54 & 39.38 & 2986.9  & 80721 \\
NGC 5053/M53 & 199.1 & 18.3 & 5245.3 & 80733 \\
NGC 5053/M53 & 198.3 & 17.5 & 7200.4 & 80863 \\
M 5  & 229.64 & 2.58  & 10112.7  & 80734\\
M 13 & 250.42 & 36.96  & 7242.2 & 80735\\
M 44 & 130.1166 & 19.4692 & 1200.1 & 80719 \\
M 92 & 259.28 & 43.64 & 11076.0 & 80736 \\
\hline
GD 1 & 128.5 & 0.8 &  4225.2 & 80722 \\
GD 1 & 132.7 & 9.93 & 655.9 & 80724 \\
GD 1 & 132.7 & 9.93 & 2400.2 & 80648 \\
GD 1  & 163.736 & 47.858 & 8488.8  & 80658 \\
GD 1  & 163.74 & 47.86 & 9754.7 & 80729 \\
GD 1  & 173.55 & 52.86  & 7153.1 & 80730 \\
Orphan & 159.08 & 7.5 & 2576.3 & 80727 \\
\hline
Ursa Major 2  & 132.87 & 63.73  & 4239.3 & 80726 \\
Ursa Major 2  & 132.87 & 63.73  & 1200.1 & 80720 \\
Sextans 1  & 154.1 & -1.375  & 1431.2 & 80614\\
Sextans 1 & 153.26 & -1.11 & 3600.1 & 80737 \\
Draco 1  & 260.07 & 58.42  & 7720.7 & 80738 \\
Draco 1  & 260.07 & 57.42  & 6100.3 & 80862 \\
\hline
M31 & 10.17 & 41.38 & 2700.1 & 80715 \\
M33 & 24.027 & 31.39  & 3600.1 & 80615 \\
\hline
Blank &  98.0063 & -4.8356   \tablefootnote{The field was designed to point at the Rosette Nebula, but was mistakenly centred at a negative declination of $\delta=-4.8$ instead of $\delta=4.8$.}
 & 1800.3 & 80718\\
Blank &  192.86 & 27.13 & 6663.5 & 80731\\
Blank &  198.04 & 0.0 &  5914.9  & 80732 \\
\hline
\end{tabular}
\label{tab:mws_sv1_areas}
\end{table}

\section{MWS Processing}
\label{sec:processing}

\subsection{DESI data}

The MWS VAC described in this paper is based on the processing of coadded spectra of DESI targets grouped by sky area \citep[HEALPix pixels with {\tt nside} = 64;][]{gorski2005}, survey and program. This VAC does not include the processing of spectra from individual exposures or spectra coadded across surveys and programs. Importantly, within a given survey and program, if an object was observed multiple times on different nights or within a different pointing, all the spectra of that object were coadded. This means that, even within one HEALPix pixel, it is possible to have targets with very different total exposure times and numbers of exposures. The details of the coaddition process are provided in Section 4.1.1 of  \citet{guy22a}.

The processing steps described below are all executed after the data reduction and flux calibration by the main DESI pipeline \citep[][]{guy22a}, as well as spectral classification and redshift determination by {\tt Redrock} (Bailey et al. in prep).

\subsection{Stellar subset}

The MWS pipelines are run on a subset of the targets observed in DESI to avoid fitting stellar models to spectra of quasars or galaxies, which are much more numerous among DESI targets. 
This VAC release is based on the subset of DESI spectra that satisfy at least one of the following criteria:

\begin{itemize}
    \item The object was targeted as an MWS object (i.e.\ has the {\tt MWS\_ANY} bit set in the {\tt DESI\_TARGET} ). 
    \item The object was targeted as a secondary target (i.e.\ has the {\tt SCND\_ANY} bit set in {\tt DESI\_TARGET}).
    \item The object was targeted as a standard star (i.e.\ it has one of the {STD\_} bits set in {\tt DESI\_TARGET} ).
    \item The object was classified by {\tt Redrock} as a star.
    \item The object has a {\tt Redrock} redshift between $-1500\,\mathrm{km\,s^{-1}}$ and $1500\,\mathrm{km\,s^{-1}}$ (irrespective of spectral type)\footnote{This selection is expected to bring significant number of non-stellar objects, but ensures that we do not miss possible stars with high velocities.}.
\end{itemize}

These criteria likely include the vast majority of stars observed by DESI, only potentially missing stars targeted through one of the cosmological DESI surveys and incorrectly classified by {\tt RedRock} as something other than a star.
We also require that the object's spectrum has a median (across all the pixels) signal-to-noise ratio per pixel of at least two in one of the B, R and Z arms and that the object has a good quality spectrum according to the pipeline flag {\tt COADD\_FIBERSTATUS=0}.
    
   \subsection{RVS pipeline}
   
    The first stellar pipeline that  DESI MWS runs is the {\tt RVSpecFit} code\footnote{\url{https://github.com/segasai/rvspecfit}}, or RVS, which performs the simultaneous least-squares modelling of the blue, red and infrared parts of the DESI spectra by interpolating templates from PHOENIX spectral library. The code is based on the approach presented in \citet{Koposov2011, li2019} and is described in more detail in \citet[][]{Cooper23}. RVS provides maximum likelihood estimates of stellar parameters surface gravity \logg, effective temperature \teff, metallicity  \feh, alpha abundance \alphafe, radial velocity and stellar rotation velocity ($v \sin i$).  The uncertainties for the stellar parameters come from the Hessian matrix of the log-likelihood function, and the radial velocity uncertainties are determined from its 1-D posterior.
    The RVS modelling only relies on DESI spectral information; no priors from photometry or {\it Gaia} astrometry are used.
    
   \subsection{SP pipeline} \label{spsteps}
   
   The second stellar pipeline for MWS is the SP\footnote{An abbreviation of Stellar Parameters.} pipeline, which uses the {\tt FERRE} code\footnote{\url{http://github.com/callendeprieto/ferre}; see also \url{http://github.com/callendeprieto/piferre} for DESI-specific supporting code} to process the same set of stars as RVS.  SP works with spectra that are shifted to zero velocity, according to the velocity measurement obtained from RVS, and are continuum-normalised by dividing the observed spectra by a running mean filter with a width of 500 pixels. 
   
   The SP processing consists of two steps. In the first step, the code minimises the $\chi$-square between the model and observations in the entire DESI spectral range to find the main atmospheric parameters: \teff, \logg, microturbulence $\xi$, metallicity \feh, and alpha-element abundance \alphafe\footnote{For white dwarfs, only \teff and \logg are measured}.  In the second step, the code holds the atmospheric parameters fixed and infers individual elemental abundances for C, Mg, Ca, and Fe\footnote{When the abundances of C and Fe are determined, the [Fe/H]  of the model is allowed to change. When the abundances of Mg and Ca are determined, the $\alphafe$ of the model is allowed to change.} by evaluating the $\chi$-square in regions dominated by atomic and molecular transitions associated with each element. This methodology is similar to that used in the APOGEE survey \citet{2016AJ....151..144G}.
   {\tt FERRE} was first used in \citet{Allende2006} and more details are given 
   in \citet{Cooper23}.
   
\section{Data products}
\label{sec:dataproducts}
Here we describe the data products included in this publicly available VAC\footnote{\url{https://data.desi.lbl.gov/doc/releases/edr/vac/mws/}}. 
These include:
\begin{itemize}
    \item The outputs of the RVS pipeline fitted to coadded spectra grouped by HEALPix pixels on the sky, survey and program (3944 individual pixel, survey, program combinations).
    \item The outputs of the SP pipeline fitted to the same data.
\end{itemize}

For ease of use, we also combine the tables produced by RVS and SP pipeline across HEALPix pixels, making 12 distinct SP and RVS catalogues corresponding to different survey and program combinations listed in Table~\ref{tab:survey_program}. Finally, we also provide the main combined catalogue, which includes RVS and SP measurements for all stars observed across all different surveys and programs. This is the data product that will be of most interest to users.

\change{We add that the original spectra that were used for the fits  with all the associated information are available as part of the main DESI EDR data release.}

The data model of all the products in this release is available at \url{https://desi-mws-edr-datamodel.readthedocs.io/}. The following is a brief overview.

\subsection{RVS data products}
\label{sec:rvs_contents}
   
    We start by describing the individual RVS outputs of fitting results on coadded DESI spectra.
    The output of RVS (for a given survey, program and HEALPix pixel) consists of two files:
    an {\tt rvtab} file that tabulates measurements by {\tt RVSpecFit} and a {\tt rvmod} file that contains the best-fit model for each spectrum.
    
    The {\tt rvtab} file is a FITS table with several extensions: {\tt RVTAB}, {\tt FIBERMAP}, {\tt SCORES}, {\tt EXP\_FIBERMAP}.
    The {\tt RVTAB} extension contains the main measurements made by {\tt RVSpecFit} and basic quantities that identify the source. All other extensions are copied directly from DESI data products described in \citet{guy22a}\footnote{The {\tt FIBERMAP} extension contains the targeting information, including DESI Legacy Imaging Survey fluxes, {\it Gaia} DR2 parameters and targeting bitmasks. The {\tt SCORES} extension contains quality statistics for the spectra.}.
    The full list of columns in the {\tt RVTAB} extension is given in Table~\ref{tab:rvtab}. The data in these columns are as follows:
    \begin{itemize}
    \item  {\tt VRAD}, {\tt VRAD\_ERR}, {\tt VRAD\_SKEW}, {\tt VRAD\_KURT} contain radial velocity measurements in the heliocentric frame, together with the estimate of the uncertainty, skewness and kurtosis of the radial velocity posterior.
    \item {\tt VSINI} the stellar rotation velocity $v\sin i$. In this release, the rotation velocity measurements have been affected by a software bug and should not be used.
    \item {\tt LOGG}, {\tt TEFF}, {\tt ALPHAFE}, {\tt FEH}, {\tt LOGG\_ERR}, {\tt TEFF\_ERR}, {\tt ALPHAFE\_ERR}, {\tt FEH\_ERR} are stellar parameters (surface gravity, effective temperature, $\alphafe$, $\feh$) and their uncertainty estimates from Hessian matrices of the likelihood function.
    \item {\tt NEXP} is a deprecated column for the number of individual spectra used in the fit.
    \item {\tt CHISQ\_B}, {\tt CHISQ\_R}, {\tt CHISQ\_Z}, {\tt CHISQ\_TOT} are the $\chi^2$ of the best-fit models in the blue, red, infrared arms and combined across all the arms, respectively. Note that these $\chi^2$ are not divided by the number of degrees of freedom.
    \item  {\tt CHISQ\_C\_B}, {\tt CHISQ\_C\_R}, {\tt CHISQ\_C\_Z}, {\tt CHISQ\_C\_TOT } are the $\chi^2$ of the best-fit continuum-only model, i.e. model using a smooth featureless stellar spectrum template.
    \item {\tt RVS\_WARN} is an integer bitmask set to zero if there are no warnings during the RVS fit. The first bit (or a bit number zero) of the bitmask is set if the {\tt CHISQ\_TOT} - {\tt CHISQ\_C\_TOT }$<$ 50, which means that the stellar fit provides only a small improvement over the continuum-only fit. The second bit of {\tt RVS\_WARN}
    is set to one if the best-fit velocity is within 5 \kms of the radial velocity grid edges ($\pm$1500 \kms); the third bit is set to one if the radial velocity error value is larger than 100 \kms. The remaining bits of the bitmask are unused in this VAC.
    \item {\tt REF\_CAT} and {\tt REF\_ID} are the external catalogue name and the source identifier if a catalogue other than the DESI Legacy Imaging Survey was used for targeting. {\tt REF\_CAT} can be T2 for Tycho-2, G2 for \gaia DR2, L3 for the Siena Galaxy Atlas (SGA) 2020 \citep{moustakas2023}\footnote{\url{https://www.legacysurvey.org/sga/sga2020/}}, or F1 (for sources prepared for DESI first light observations). For most \gaia sources, the {\tt REF\_ID} is the \gaia DR2 {\tt source\_id} and {\tt REF\_CAT} is G2\footnote{We do not recommend using the {\tt REF\_ID} from this table, but instead recommend using the cross-match with \gaia DR3 provided in the combined main MWS catalogue (Section \ref{sec:combined_catalogs}). This is because {\tt REF\_ID} is incorrect for some secondary targets due to a software bug with the fibre assignment code.}. 
    \item {\tt TARGET\_RA}, {\tt TARGET\_DEC} are the coordinates of objects.
    \item {\tt TARGETID} is the unique DESI target identifier\footnote{We remark that different objects observed by DESI are guaranteed to have different {\tt TARGETID}, but it is possible for the same source to have more than one {\tt TARGETID}.}.
    \item {\tt SN\_B, SN\_R, SN\_Z} are the median signal-to-noise ratios in the blue, red, and infrared arms of the spectrum of the object.
    \item {\tt SUCCESS} is a boolean flag that is set to true when the spectrum has no warnings from {\tt RVSpecFit} (i.e. {\tt RVS\_WARN}=0).
    \end{itemize}

The {\tt rvmod} FITS files contain the best-fit stellar model spectra corresponding to the individual objects in the {\tt rvtab} file. The files contain three {\tt B\_MODEL}, {\tt R\_MODEL}, {\tt Z\_MODEL} extensions for the best-fit model in each wavelength arm. The {\tt rvmod} files also include {\tt B\_WAVELENGTH}, {\tt R\_WAVELENGTH} and {\tt Z\_WAVELENGTH} extensions with the wavelength arrays for the different arms (copied from the original DESI data products).

For each combination of survey and program, in addition to catalogues of parameters measured from spectra grouped by HEALPix pixels on the sky, we also provide a combined catalogue of all objects observed that given survey and program. For example, {\tt rvpix-sv2-backup.fits} includes one row for every target in the `backup' program of the SV2 survey. These catalogues are concatenated {\tt rvtab} files from different HEALPix pixels and thus have the same structure as the individual {\tt rvtab} tables.

\subsection{SP data products}
\label{sec:sp_contents}

As for RVS, the SP output consists of two files per input spectral file with coadded DESI spectra.  One output is the {\tt sptab} file, which is the table of measurements from {\tt FERRE}, and the other is the {\tt spmod} file, which contains the best-fit model for each spectrum. The {\tt spmod} files, unlike the RVS outputs, provide continuum-normalised observed spectra, along with their uncertainties, the continuum-normalised best-fit models, and the model absolute fluxes.
    
The {\tt sptab} file is a FITS table with four extensions: {\tt SPTAB}, {\tt FIBERMAP}, {\tt SCORES}, and {\tt AUX}. The {\tt SPTAB} extension contains the main measurements from {\tt FERRE} and several additional quantities helpful for cross-matching with other data products. The {\tt FIBERMAP} and {\tt SCORES} extensions are copied directly from the main DESI data products -- they are identical to the extensions with the same names in the {\tt rvtab} files. Table~\ref{tab:sptab} gives a complete description of all the fields included in the {\tt SPTAB} extension, but we provide a summary of the most relevant ones below:
    \begin{itemize}
    \item  {\tt FEH}, {\tt ALPHAFE}, {\tt LOG10MICRO}, {\tt TEFF}, and  {\tt LOGG} are the main atmospheric parameters: \feh, \alphafe, micro-turbulence, effective temperature and surface gravity. These values are also stored, in this order, in the array {\tt PARAM} (see below), and their covariance matrix is provided in {\tt COVAR}.
    \item {\tt ELEM} and {\tt ELEM\_ERR} are arrays with the Fe, Ca, C and Mg abundances and their estimated uncertainties, respectively. See Section \ref{sec:limitations} for caveats on the Ca, C and Mg abundances.
    \item {\tt CHISQ\_TOT}  is the reduced $\chi^2$ of the best-fit model for the entire (continuum-normalised) spectrum.
    \item {\tt REF\_CAT} and {\tt REF\_ID} are the source catalogue names and the corresponding identifiers in those catalogues used for targeting -- the same as in the {\tt RVTAB} extension.
    \item {\tt TARGET\_RA}, {\tt TARGET\_DEC} are the coordinates of the object.
    \item {\tt TARGETID} is the DESI target identifier.
    \item {\tt SNR\_MED} gives the median signal-to-noise of the spectrum across B, R and Z arms.
    \item {\tt PARAM} is the vector of atmospheric parameters (\feh, \alphafe, $\log_{10}$ of microturbulence velocity, \teff, \logg).
    \item {\tt COVAR} Covariance matrix for the parameter vector given in {\tt PARAM} column.
    \item {\tt SRCFILE} is the file name with the DESI spectrum that was fitted.
    \item {\tt BESTGRID} is the stellar model grid that provided the best fit to the spectrum. Possible values are 'm\_rdesi1' to 'm\_rdesi5' for Kurucz models, 'm\_rdesi6' to 'm\_rdesi9' for Koester white-dwarf models, or 's\_rdesi1' for PHOENIX models.
    \item {\tt RV\_ADOP}, {\tt RV\_ERR} are the radial velocity and its uncertainty (adopted from the RVS fits) when shifting the spectrum to the rest frame. 
    
    \end{itemize}

The {\tt AUX} extension of the {\tt sptab} files includes two string arrays: the {\tt p} array gives the names of the parameters in the order included in the {\tt PARAM} array and the {\tt COVAR} matrix, and the {\tt e} array gives the names of the chemical elements included in the {\tt ELEM} and {\tt ELEM\_ERR} arrays.

The model {\tt spmod} files are FITS files with nine extensions: {\tt B\_WAVELENGTH}, {\tt R\_WAVELENGTH} and {\tt Z\_WAVELENGTH}, which store the wavelengths arrays, the {\tt B\_MODEL}, {\tt R\_MODEL}, {\tt Z\_MODEL}, which hold the model spectra and related quantities, the {\tt FIBERMAP} and {\tt SCORES} extensions, already mentioned in the description of the {\tt sptab} files, and a {\tt FILTER} extension, which contains the concatenated arrays of weights for the {\tt B}, {\tt R} and {\tt Z} channels used for determining elemental abundances.  

The model extensions ({\tt B\_MODEL}, {\tt R\_MODEL}, and {\tt Z\_MODEL}) contain multiple columns:
\begin{itemize}
    \item The {\tt OBS} column in the tables gives the continuum-normalised observed spectrum resampled to rest-frame. 
    \item The {\tt ERR} column stores the uncertainties on the continuum normalised resampled spectrum. 
    \item The {\tt FIT} column contains the best fit continuum-normalised stellar model obtained from the first step described in \S \ref{spsteps}, corresponding to the determination of the stellar parameters. 
    \item The {\tt ABU} column is similar to the {\tt FIT} column, but for the fitting obtained in the second step (abundances determination). There is only one {\tt ABU} array but multiple elements, so the data are overwritten in the order given in the {\tt ELEM} array, chosen to ensure that values corresponding to elements with fewer spectral lines are written last. The {\tt FLX} column stores the un-normalised absolute flux model corresponding to the best-fit parameters of the star.
\end{itemize}

Similar to the RVS outputs, we provide combined catalogues of objects observed in a given survey and program, for example {\tt sppix-sv2-dark.fits}.  These are constructed by concatenating {\tt sptab} files and thus have a similar structure as the individual {\tt sptab} files, but without the {\tt AUX} extension.

\subsection{Main combined catalogue}\label{sec:combined_catalogs}

In addition to the catalogues of measurements for targets grouped by DESI survey/program and HEALPix pixel, we also provide a combined catalogue of measurements from RVS and SP pipelines, including all the surveys and programs that are part of the DESI EDR. This main catalogue is a FITS table named {\tt mwsall-pix-fuji.fits}. The main catalogue also includes additional \gaia DR3 cross-match information. 
We expect the main combined catalogue to be the most frequently used data product in the DESI MWS VAC. The catalogue contains 625,588 rows.

The combined table contains the following extensions: {\tt RVTAB}, {\tt SPTAB}, {\tt FIBERMAP}, {\tt SCORES}, {\tt GAIA}.
The {\tt RVTAB} extension includes all the columns listed in Section~\ref{sec:rvs_contents}, as well as a few additional columns:
\begin{itemize}
    \item {\tt PROGRAM} identifies the program from which the star came from (i.e. dark, bright, backup or other).
    \item {\tt SURVEY} identifies the survey from which the star came from (i.e. SV1, SV2, SV3 or CMX).
    \item {\tt HEALPIX} is the HEALPix pixel (at {\tt nside} = 64) of the star. 
    \item {\tt PRIMARY} is the flag that is set to true if this specific measurement was labelled as primary (because the same star could have been observed in different surveys and programs). The observation with the largest signal-to-noise in the R arm is selected as primary.
    \item {\tt RR\_Z} is the redshift from {\tt Redrock}.
    \item {\tt RR\_SPECTYPE} is the spectroscopic type of the object, such as {\tt STAR}, {\tt QSO}, {\tt GALAXY} as provided by {\tt Redrock}.  
\end{itemize}

The {\tt SPTAB} extension in the catalogue contains measurements from the SP pipeline as described in Section \ref{sec:sp_contents}.
 The {\tt FIBERMAP} and {\tt  SCORES} extensions store information relevant to targeting and spectral quality, respectively \citep[see][]{guy22a}. Finally, the {\tt GAIA} extension in the file contains \gaia DR3 measurements for each source in the catalogue (we provide all the columns from the {\tt gaia\_source} table). \gaia sources have been associated with DESI targets by on-sky position (using a radius of one arcsecond).

\section{Catalogue overview}
\label{sec:analysis}

In this section, we provide an overview of the contents of the MWS VAC catalogue, demonstrate how it could be used for science, and discuss some of its limitations.

\subsection{Selecting clean stellar samples}
\label{sec:clean_sample_selection}

\begin{table}
\caption{The number of objects in the value-added catalogue with different {\tt Redrock} spectroscopic types, separated according to whether the {\tt RVS\_WARN=0} condition is true or false.}
\begin{tabular}{lll}
\hline
{RR\_SPECTYPE} & 
{RVS\_WARN=0} & 
{Number}
\\
\hline
STAR & True & 511,809 \\ 
STAR & False & 4,454 \\ 
GALAXY & True & 13,911\\ 
GALAXY & False & 61,677 \\ 
QSO & True & 10,336\\ 
QSO & False & 23,401 \\
\hline
\end{tabular}
\label{tab:spectype}
\end{table}

\begin{figure}
    \centering
    \includegraphics[]{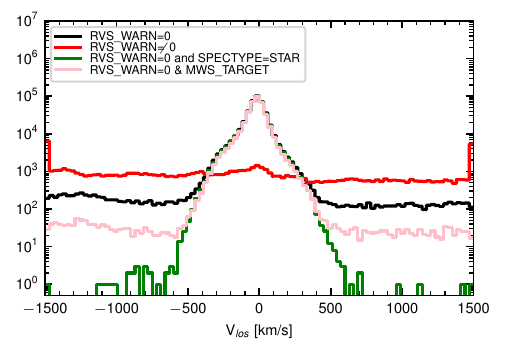}
    \caption{The distribution of line of sight velocities for different subsamples in the VAC. The distributions highlight alternative ways of selecting stellar objects from the catalogue. The black line shows the distribution of objects with {\tt RVS\_WARN=0}. These comprise mostly stars, with a few per cent contamination by non-stellar sources that produce flat tails in the distribution out to $\pm$1500\kms. The red line shows the distribution of objects with {\tt RVS\_WARN}$\neq$0. These are almost uniformly distributed in velocity, confirming that they are not stars.
    The pink curve shows the distribution of radial velocities for those objects with {\tt RVS\_WARN}=0 that were targeted for the MWS survey; the green curve shows the distribution for objects with {\tt RVS\_WARN}=0 that were classified as star by {\tt Redrock} (i.e. {\tt RR\_SPECTYPE}=STAR). The latter sample has the least contamination; these criteria should be used to select stars in the MWS VAC unless stars with $|V_\text{los}|>600$\kms are of interest.}
    \label{fig:rvhist}
\end{figure}

Because the list of criteria used to identify potential objects for fitting by the stellar pipeline (described in Section~\ref{sec:processing}) is deliberately broad 
and designed not to miss any stellar targets, the MWS catalogue contains a significant number of sources that are not stars. 

In this section, we provide several methods to identify mostly clean samples of stars using values provided in the VAC tables. These criteria are not meant to be exhaustive or optimal because different science cases may have different requirements for purity and completeness.

\begin{itemize}
    \item {\tt RVS\_WARN = 0} selects objects for which {\tt RVSpecFit} obtained a satisfactory fit to the DESI spectrum. This selects most stellar spectra, but this is not a particularly pure selection because it accepts a fraction of QSOs and galaxies. 
    
    \item {\tt RR\_SPECTYPE = STAR} selects on the {\tt Redrock} classification of an object as a star. This is a robust classification, but it is limited by the condition imposed by {\tt Redrock} that stars have radial velocities between -600  and 600 \kms. 
    
    \item  ({\tt DESI\_TARGET \& MWS\_ANY})$>0$\footnote{Where {\tt MWS\_ANY}=$2^{61}$ as specified in Table~\ref{tab:main_desi_target}.} selects objects with the \texttt{MWS\_ANY} targeting bit set, which means that the source was targeted for the Milky Way Survey. This removes a large number of non-stellar objects but also retains a significant number of QSOs that fall in the MWS \texttt{MAIN-BLUE} and \texttt{BROAD} selections. 
\end{itemize}

In Table~\ref{tab:spectype}, we provide the number of objects in the catalogue with different spectroscopic classifications from {\tt Redrock}, depending on whether {\tt RVS\_WARN = 0} is true or not. The table shows that the number of non-stellar objects in the MWS VAC EDR catalogue is $\sim$ 20 per cent.

Figure~\ref{fig:rvhist} shows the distribution of radial velocities as measured by RVS and reported in the MWS VAC. We show separately the distributions for objects with {\tt RVS\_WARN}$=0$ and {\tt RVS\_WARN} $\neq 0$ (black and red lines respectively), illustrating that objects with {\tt RVS\_WARN} $\neq 0$ do not concentrate at $V_\mathrm{los}=0 \kms$. This indicates that objects with {\tt RVS\_WARN} $\neq 0$ are either not stars or unlikely cases of a bad fit by the RVS pipeline. The figure also shows that an additional constraint on the spectroscopic type from {\tt Redrock} (green line) or on selection as an MWS target (pink line) results in a velocity distribution concentrated mostly within $|V_\mathrm{los}|<600$\kms, implying a cleaner stellar selection.

For the rest of this paper, we will use the {\tt RVS\_WARN=0} and {\tt RR\_SPECTYPE=STAR} selection to identify stellar targets as well as the \texttt{PRIMARY=True} flag to exclude duplicate observations. This leaves 421,403 individual sources out of 625,588 in the combined catalogue. Additional cuts will be noted where relevant.

A few further selection cuts are recommended for restricting the sample to the highest-quality measurements. These are described in Appendix \ref{sec:sp_cuts} for SP and in Appendix \ref{sec:rvs_cuts} for RVS.

\begin{figure}
\includegraphics[]{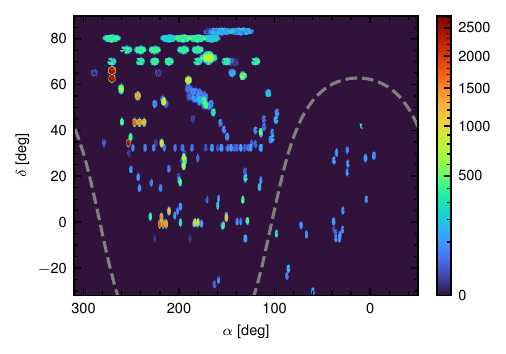}
\caption{The density on the sky of stellar targets in the value-added catalogue. The density is shown in units of stars per square degree.  The dashed line shows the Galactic plane. Individual DESI pointings are easily discernible on the map as 3-degree diameter circles.}
\label{fig:skymap}
\end{figure}

\subsection{Spatial and colour-magnitude distribution}
\label{sec:spatial_magnitude}
\begin{figure*}
\includegraphics{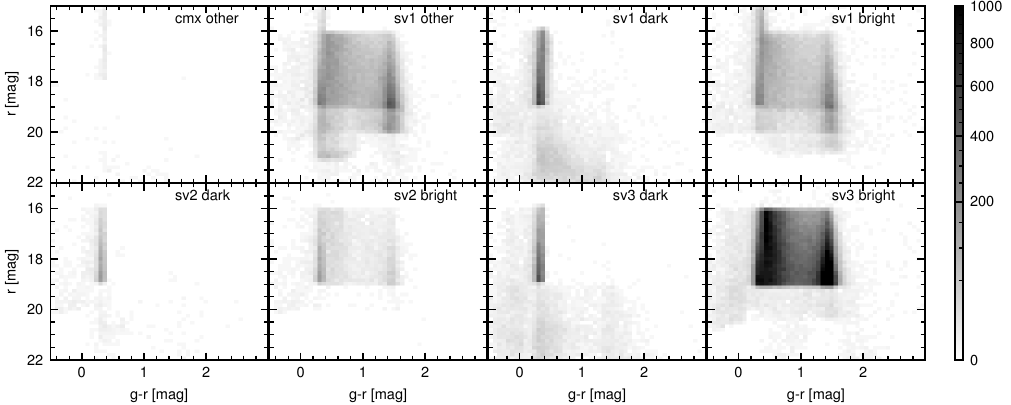}
\caption{Hess diagrams of stellar targets in the VAC observed in different surveys and programs. The bright stars with $g-r\sim 0.4$ are those used by the DESI survey as flux calibration standards. As the main MWS survey targets are only observed in bright time,  the main $16<r<19$ selection is only seen in the bright SV programs (as well as in SV1/other, which was used for some dedicated MWS tiles). In the dark program, stellar targets are observed either as standards, secondary targets or stellar contaminants from other cosmological target selections.
\label{fig:cmd}}
\end{figure*}

\begin{figure}
    \centering
    \includegraphics[width=\columnwidth]{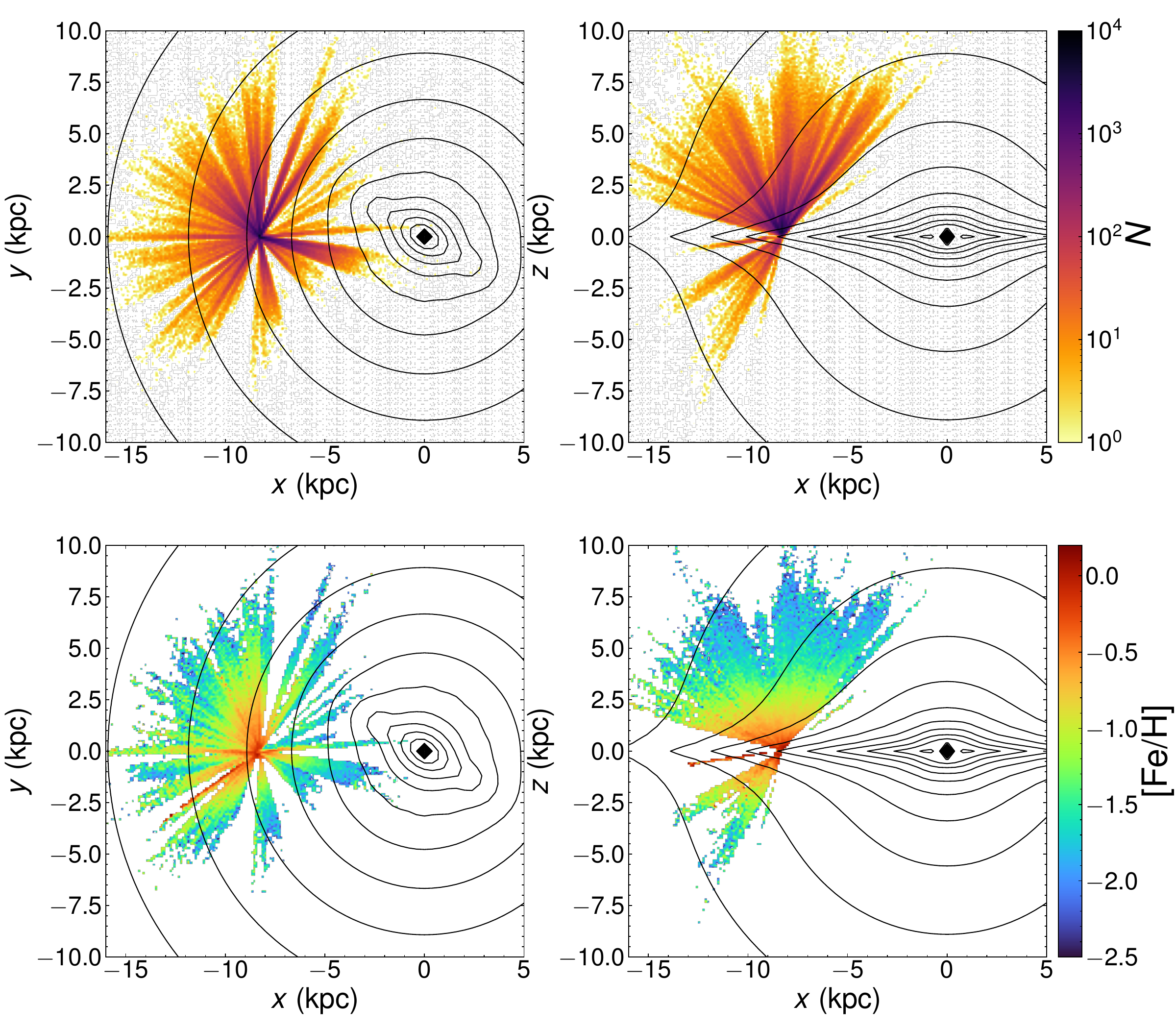}
    \caption{Spatial distribution of stars from the DESI EDR value-added catalogue in Galactocentric coordinates $x-y$ (left) and $x-z$ (right), in $200\times 200$ bins. The panels are colour-coded by the numbers of stars per bin (top) and mean $\feh$ (bottom) from the RVS pipeline. The solid contours show iso-potentials of the barred Milky Way model of Hunter, Sormani+ 2023 (in prep.). The distribution is based on distances from \citet{Bailer-Jones2021} and thus shows only the nearby part of the DESI sample.}
    \label{fig:xyz_maps}
\end{figure}

The distribution of VAC stars on the sky is shown in Figure~\ref{fig:skymap}. The colour on this map shows the density of observed targets, which varies from a few hundred to a few thousand per square degree, depending on which surveys (SV1, SV2 or SV3) and programs covered a particular area. The density is highest in dedicated MWS fields (described in Section~\ref{sec:dedicated_fields}) and SV3 areas, which have high completeness for all target classes. In areas of the sky that were not observed by the bright SV programs, the target density is typically small. Stars in these fields were only observed as flux standards or secondary targets or as contaminants to other DESI target selections.

The colour-magnitude diagram of stars from all the surveys and programs included in the VAC is shown in Figure~\ref{fig:cmd}. This figure highlights several aspects of target selection discussed in Section~\ref{sec:dataset}. A common feature of all these colour-magnitude distributions is the area populated by DESI flux calibration standard stars, with $g-r\sim 0.4$ and $15 \lesssim r\lesssim 19$. These targets are included in both dark and bright observations. In the bright program observations, the colour-magnitude distribution shows the main MWS selection in the magnitude range $16<r<19$ and fainter extensions to $r<20$ explored during SV.

Figure~\ref{fig:xyz_maps} illustrates the distribution of stars from the VAC in Galactocentric coordinates using
photo-geometric distance estimates from \citet{Bailer-Jones2021}\footnote{We assume the distance to the Galactic centre $R_0=8.277$\,kpc \citep[][]{Gravity_2022}, and the solar height with respect to the Galactic plane $z_0 = 20.8$~pc \citep{Bennett_2019}.}. The top two panels show the 2-dimensional number density of stars in Galactocentric $x$ vs $y$ and $x$ vs $z$.  The distribution shown in this figure is only valid within $\sim 10 $\,kpc of the Sun because the distances from \citet{Bailer-Jones2021} are not robust enough for more distant targets. The bottom row of the figure shows the distribution of stars in the same coordinates, but now colour-coded by mean [Fe/H] from the RVS pipeline. The solid contours show the iso-potentials of the barred Milky Way model of Hunt, Sourmani et al. (in prep.), built with the Agama library \citep[see][]{Vasiliev2019} -- see also \cite{Sormani2022}. The figure illustrates the Galactic volume probed by the majority of the DESI EDR stellar sample, with most stars at high galactic latitudes.  Although a significant part of the DESI EDR sample shown in the figure is within a Galactocentric distance of $\sim 15$\,kpc, a large number of stellar targets in the VAC sample of the Milky Way stellar halo out to distances of $\sim 100$ kpc. This is not visible in the Figure due to the distance estimator used (better distance estimators are currently being developed). The bottom panels of the figure also illustrate that we recover the expected decrease in mean [Fe/H] further from the disk mid-plane ($|z|$), with high-$|z|$ halo stars typically having $\feh \lesssim -1$.

\subsection{Stellar parameter distribution}
\label{sec:stellar_param_distribution}
\begin{figure}

\includegraphics{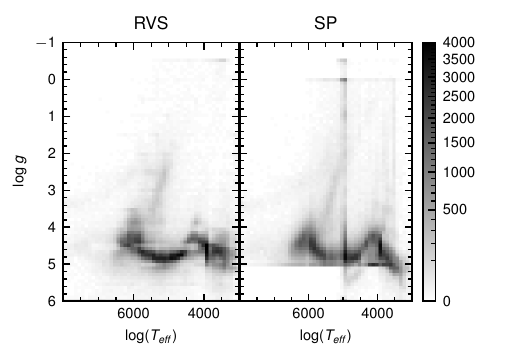}
\caption{The distribution of surface gravity and effective temperatures for stars in the MWS VAC as determined by the RVS and SP pipelines. The parameter range shown here is smaller than the actual range of measured parameters; the RVS maximum temperature extends to $15,000$~K and SP to $40,000$~K. The SP pipeline also models white dwarfs, which can have \logg up to 9.5 and are not shown here.} 
\label{fig:params}
\end{figure}

All of the stars in the MWS VAC have stellar parameter measurements from both the SP and RVS pipelines. We provide measurements of \logg, \teff and \feh (as well as other parameters described in Sections~\ref{sec:rvs_contents} and \ref{sec:sp_contents}).

In Figure~\ref{fig:params}, we show the distributions of surface gravity and effective temperature (Kiel diagrams) for stars in the VAC, as measured by the RVS (left) and SP pipelines (right). The distributions are similar and show features that are expected, such as the red giant branch (extending from $\logg=4$, $\teff=5500$\,K to $\logg=1$, $\teff=4000$\,K), the blue horizontal branch (\logg$\sim$3, $\teff>6,500$\,K) and the main sequence (\logg<4, \teff<6,000\,K). The shape of the main sequence below the turn-off, however, is not very realistic and is likely dominated by systematic effects from the stellar atmospheric grids used in our pipelines. Both pipelines show clustering around grid points in \logg and \teff. In addition, the SP pipeline also shows problematic clustering at $\teff=5000$\,K and $\logg=5$. Some of these SP artefacts mostly disappear when using stricter selections described in Appendix~\ref{sec:sp_cuts}.

\change{In Figure~\ref{fig:fehdistr} we show the distribution of iron abundances for all the unique DESI stars. We display the distribution for abundances from  both RVS and SP pipelines. The metallicity distribution  function (MDF) shows that the majority of stars is in the thick disk with $\feh \approx - 0.5$, but with about 25\% of stars having metallicities below $-1$. The metallicity distribution for metal-poor stars shows the peak at $\feh \approx -1.2$ which would be the GSE structure \citet{belokurov2018,helmi2018}. At low metallicities the MDF extends like a power-law down to $\feh=-3.5$ as expected for the Milky Way stellar halo \citep{youakim2020}. The metallicity distributions from RVS and SP do agree, although due to issues highlighted earlier, we see some discrepancies for very metal-rich stars, as well as accumulation at the edges near $\feh=-4$ for RVS and $\feh=-3.5$ for SP.}

\begin{figure}
\includegraphics{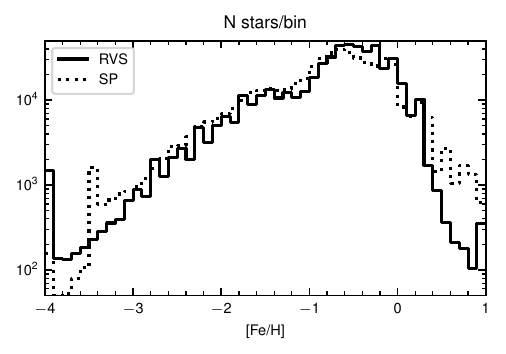}
\caption{\change{The distribution of iron abundances for stars in the MWS VAC as determined by the RVS and SP pipelines. The distribution shows the peak associated with thick disk stars at \feh=-0.5, the peak associated with GSE at $\feh=-1.2$ and metal-poor stellar halo at lower metallicities. The distributions are shown with the bin size of 0.1 dex. Due to gridding issue associated with the RVS measurements, would the plot be made with much smaller bins (i.e. 0.01 dex), we would see peaks at location of template grid with the step of 0.25 dex.}} 
\label{fig:fehdistr}
\end{figure}

\subsection{Parameter validation with APOGEE}
\label{sec:apogee_systematics}
\begin{figure}
    \centering
    \includegraphics[]{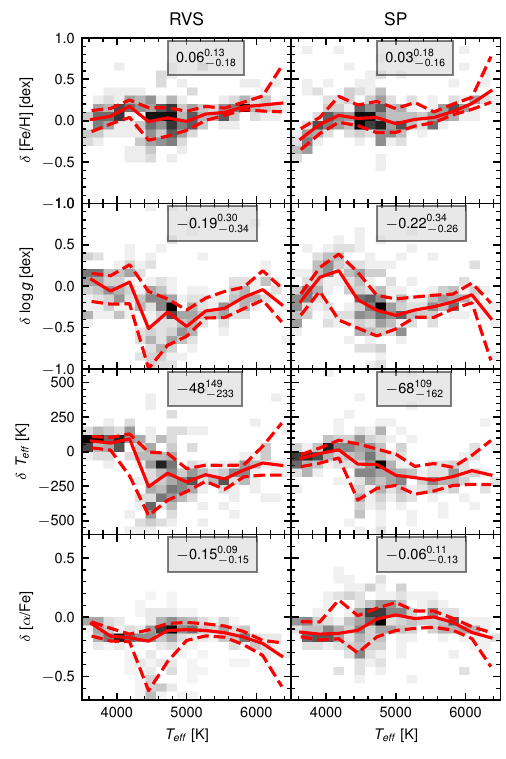}
    \caption{Comparison of RVS and SP parameters to APOGEE measurements is shown as a function of APOGEE effective temperature. Each panel shows in
     greyscale shows the 2D histograms of $\delta X = X_{DESI} - X_{APOGEE}$ (where X is $[Fe/H]$ for the top panel, $\log g$ for the second panel $T_{\rm eff}$ for the third panel and $[\alpha/Fe]$ for the bottom panel) vs effective temperature. Solid curves show the median residual with respect to APOGEE, and dashed curves show the 16th and 84th percentiles of the distribution. Inset boxes report the median value of the offset with respect to APOGEE for the whole sample.}
    \label{fig:apogee_comparison}
\end{figure}

Here, we compare our stellar parameter measurements with the parameters from the APOGEE survey in more detail than \citet{Cooper23}. The APOGEE results are of high quality (high-resolution and high signal-to-noise) and based on $H$-band spectroscopy. \citet{2018AJ....156..126J} has shown good agreement between the APOGEE results and high-resolution optical studies in the literature, although systematic differences are apparent for some elements.
We specifically take the APOGEE DR17 measurements from \citet{apogee_dr17} and cross-match them with the MWS VAC. In the APOGEE catalogue, we exclude measurements with the {\tt STAR\_BAD} flag set. In the MWS VAC, we use the selection given in Section~\ref{sec:clean_sample_selection}. After these cuts, we have 507 stars in common between APOGEE and the MWS VAC.

Figure~\ref{fig:apogee_comparison} shows the residuals between RVS and SP measurements of \logg, \teff, \alphafe and \feh and the corresponding measurements from APOGEE. The residuals are plotted as a function of the effective temperature from APOGEE. We also show the median (solid line) and the 16/84 percentiles of the residuals (dashed lines). The annotation in each panel gives the median and 16/84 percentiles of the residuals for the whole DESI$\times$APOGEE dataset.

The figure shows that, for both RVS and SP, the typical scatter in [Fe/H] with respect to the high-resolution APOGEE measurements is $\sim$ 0.15-0.2\,dex with a small bias. There are no significant trends of [Fe/H] with temperature, but the metallicities seem more robust for warmer stars with \teff$>4800$~K. The surface gravity measurements show strong systematic trends with temperature. At \teff $\sim4500$~K the RVS \logg deviation reaches $-0.5$, and at a similar temperature, the SP systematics in \logg switch from positive to negative bias with respect to APOGEE. The effective temperatures for RVS and SP show good agreement with APOGEE for cool stars but show $\sim$ 200\,K bias for hotter stars. With \alphafe the typical agreement with APOGEE for both RVS and SP is around 0.1 dex, although the dynamic range of \alphafe in the DESI/APOGEE overlap sample is quite small ($\sim $  0.4\,dex). The \alphafe measurements from RVS seem more biased than SP and there is a subset of stars with large \alphafe deviation at $\teff=4000$~K\footnote{This particular feature was caused by a few problematic stellar atmospheres in the PHOENIX grid. Those stellar atmospheres have since been removed, and the problematic feature disappears in more recent reductions of the DESI data.}

Although Figure~\ref{fig:apogee_comparison} shows systematic trends with temperature, the trends with surface gravity are also important. For example, there is an indication that PHOENIX-based measurements are biased for \logg$<2$. However, since the EDR dataset does not overlap enough with high-resolution surveys like APOGEE to investigate systematics as a function of multiple variables, we postpone a more detailed analysis of these systematics to future data releases.

\change{The comparison of abundance and stellar parameter measurements in APOGEE presented here is mostly demonstrating the systematic offsets of DESI measured parameters as opposed to random errors. As shown in \citet{Cooper23} (Figure 14) the \feh uncertainty is a strong function of magnitude and colour, reaching values of $\sim $ 0.3 dex for blue stars at the faint $r=19$ limit.}

\begin{figure}
    \centering
    \includegraphics[]{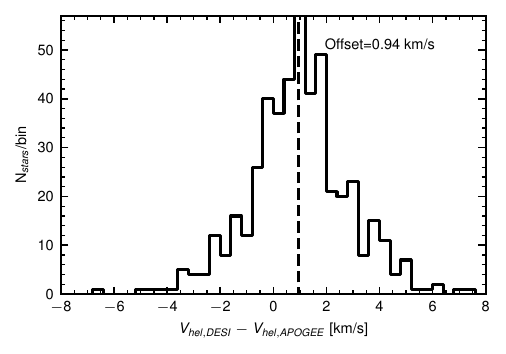}
    \caption{The histogram of differences between DESI RVS and APOGEE radial velocity measurements. The distribution shows a small bias, such that velocities from DESI are $\sim 0.93$\,\kms larger than those from APOGEE.}
    \label{fig:apogee_rv_comparison}
\end{figure}

\subsubsection{Radial velocity comparison with APOGEE}
\label{sec:rv_comparison}

In the overview paper, we analysed the overall quality of radial velocities and radial velocity errors (Figure~11 in \citet{Cooper23}). The analysis of repeated observations showed that the velocity errors provided by {\tt RVSpecFit} correctly characterise the velocity uncertainties if we take into account the systematic error floor of 0.9\kms. We do not repeat this analysis here because it should also apply to the data in the EDR. However, it is worthwhile to check for any systematic velocity offset with respect to other datasets.

In Figure~\ref{fig:apogee_rv_comparison}, we compare the distribution of radial velocity differences against APOGEE using the same sample as our investigation of the stellar parameters above. This shows that DESI radial velocities have a small bias of $0.93$\kms (i.e. DESI radial velocities are larger than APOGEE velocities by 0.93 \kms). We do not apply this correction in the VAC, so users should consider subtracting 0.93\kms from DESI velocities to bring them to the same frame as APOGEE.

\subsection{Chemo-dynamic studies with DESI stars}
\label{sec:chemodynamics}

\begin{figure}
    \centering
    \includegraphics[width=\columnwidth]{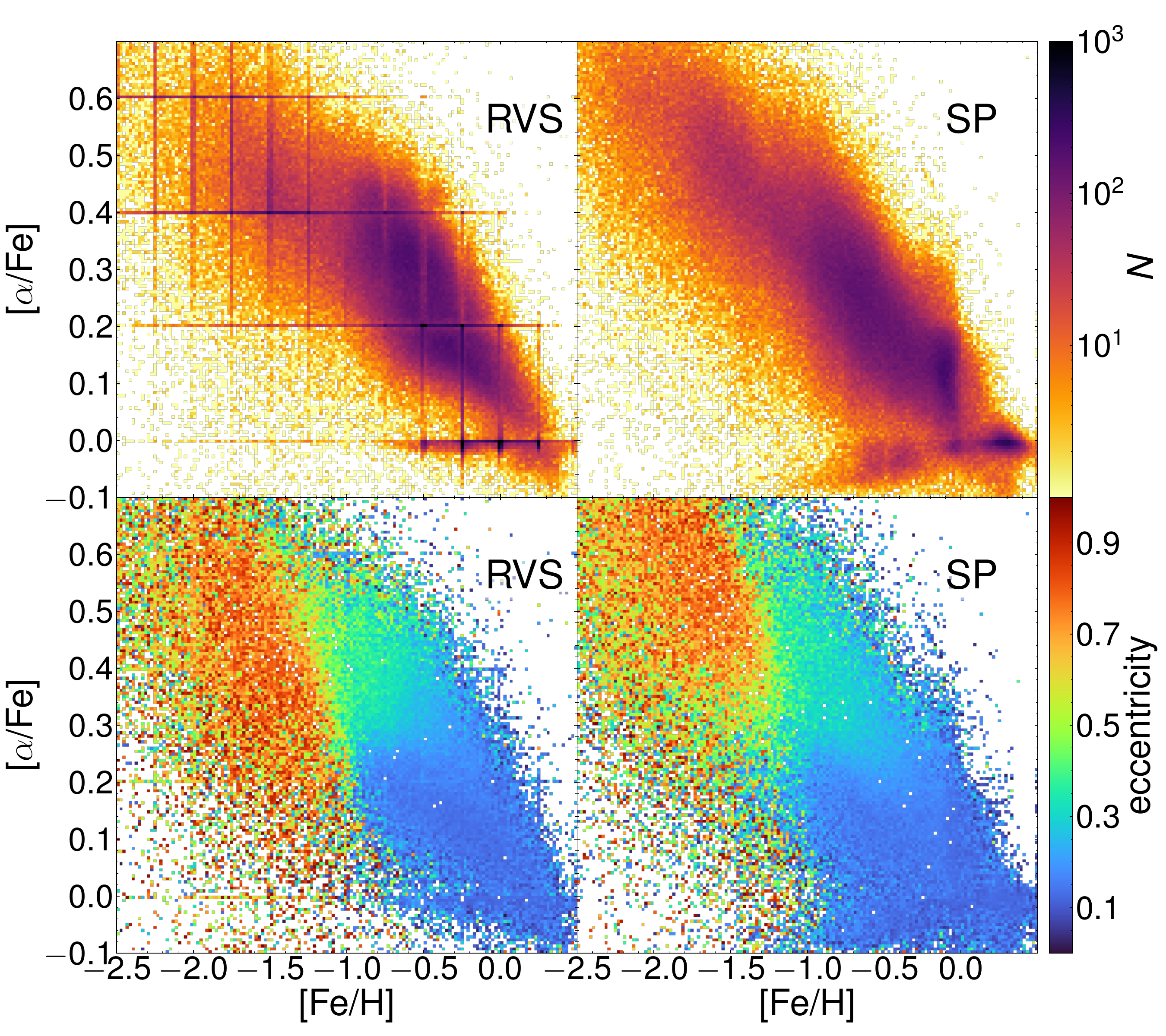}
    \caption{Chemical map ([Fe/H]-[$\alpha$/Fe]) from the RVS (left) and SP (right) pipelines, in $150\times 150$ bins. The panels are colour-coded by the number of stars per bin (top) and by the eccentricity (bottom) $(r_\mathrm{apo} - r_\mathrm{per})/(r_\mathrm{apo} + r_\mathrm{per})$, where $r_\mathrm{apo}$ and $r_\mathrm{per}$ are the apo- and pericenter radii. Orbits were integrated for $10$ Gyr in the potential model of \citet{McMillan2017}, using photo-geometric distances from \citet{Bailer-Jones2021}.}
    \label{fig:feh_afe_hist_rvs_sp}
\end{figure}

Here, we look at the spectroscopic abundance measurements in combination with the orbital properties of stars to demonstrate the power of data included in the MWS VAC for chemical analyses of the Milky Way disc and stellar halo \citep[similar to e.g.][]{mackareth2019}.
Figure~\ref{fig:feh_afe_hist_rvs_sp} shows the \feh vs \alphafe distributions for stars in the VAC.  The top panels show the density of stars in \feh vs.\ \alphafe space, while in the bottom panels, we colour each pixel by the average orbital eccentricity of stars in those pixels. The left and right panels show measurements from the RVS and SP pipelines, respectively.

Looking at the distribution of abundances in the top left panel of the figure, we notice very clear gridding effects in RVS parameters, with steps of 0.25\, dex in \feh and 0.2\,dex in \alphafe\footnote{The gridding in measured parameters is caused by the use of multi-linear interpolation over a rectangular grid in the space of parameters. This interpolation is continuous but does not produce continuous first derivatives, leading to clumping at the edges of interpolation grid cells. This will be addressed in future DESI MWS data releases.}. Despite this gridding, the overall distribution closely resembles that obtained by higher resolution spectroscopic surveys \citep[e.g. APOGEE -- see][]{apogee_dr17}, with the region $\mathrm{[Fe/H]}<-1$ being typically populated by halo stars and the region $\mathrm{[Fe/H]}>-1$ being typically split into an [$\alpha$/Fe]-rich track ($\mathrm{[\alpha/Fe]}\gtrsim 0.2$) and an [$\alpha$/Fe]-poor track ($\mathrm{[\alpha/Fe]}\lesssim 0.2$). The \alphafe-poor and \alphafe-rich sequences can be seen in the figure \change {separated at \alphafe of 0.15 at \feh=-0.5. The halo stars are seen as less dense part of the distribution at $\feh<-1$ and $\alphafe\sim 0.5$.} The distribution of SP abundances (top right panel) does not show gridding effects but resembles less closely the halo/\alphafe-rich/\alphafe-poor distribution of APOGEE. 
We also note a clustering of stars at \feh=0 and \alphafe $\sim 0$ that is not physical.


The bottom panels of Figure~\ref{fig:feh_afe_hist_rvs_sp} show the same distribution as the top panels but are colour-coded by the average eccentricity of orbits for these stars. To compute orbits, we use \gaia DR3 astrometry and the photo-geometric distances from \citet{Bailer-Jones2021} (without applying any quality cuts on the distances). Orbits were integrated for 10 Gyr in the Milky Way potential of \cite{McMillan2017} with the same parameters for the position of the Sun in the Galaxy as in Section~\ref{sec:spatial_magnitude} and assuming $U_\odot = 9.3$ km/s, $V_\odot + V_c(R_0)=251.5$ km/s and $W_\odot=8.59$ km/s, from \citep{Reid_2020, Gravity_2022}.

  Overall, both the RVS and SP chemical maps show the same chemo-dynamical correlations observed in other spectroscopic surveys in combination with \gaia: intermediate metallicity halo stars dominated by the Gaia-Enceladus-Sausage merger \citep{belokurov2018,helmi2018} typically have eccentricities $\gtrsim 0.8$, while the \alphafe-rich (\alphafe -poor) stars have intermediate (low) eccentricities, in close connection to the thick and thin disc, respectively -- see Figure~\ref{fig:xyz_maps}. \change{
    While in this data we may also expect to see separation of the GSE from other parts of the accreted halo \citet{Feuillet2021,hasselquist2021}, given the current quality of $\alphafe$ measurements we can not make such a separation. 
  We also note that in our analysis did not take into account the survey selection function which can affect Figure~\ref{fig:feh_afe_hist_rvs_sp}. One reason it may play a role is because the {\tt MWS\_MAIN\_BLUE} sample, which constitutes a substantial fraction of the catalogue is colour selected (hence is biased towards more metal-poor stars), but has no astrometric selection. The complementary red {\tt MWS\_MAIN\_RED} sample is selected using colour, but also uses astrometry to exclude more nearby stars. This will affect significantly the distribution shown on the Figure by down-weighting the metal-rich disk population.}

\subsection{Draco dwarf galaxy dedicated fields}
\label{sec:draco}
\begin{figure}
\includegraphics[]{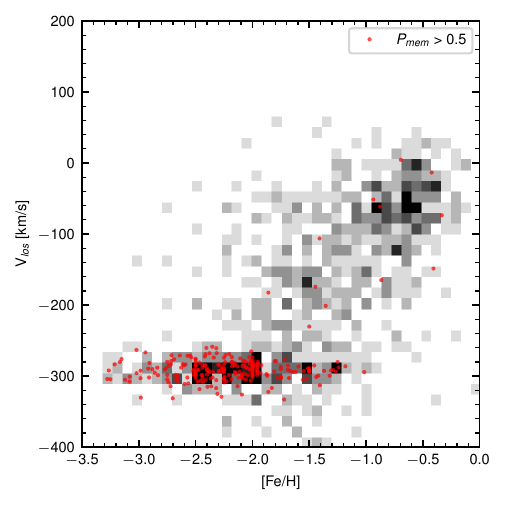}
\caption{The distribution of stars in the MWS VAC around Draco dSph in \feh and line of sight velocity. Here, we use only stars with membership probabilities based on \gaia data, computed by \citet{pace_li2022}. Greyscale shows the 2-D histogram of all the stars with membership measurement from \citet{pace_li2022}. The stars with high membership probability $P>0.5$ are shown in red. The total number of likely Draco members is $\sim$ 260 stars.}
\label{fig:draco}
\end{figure}

As mentioned in Section~\ref{sec:dataset} and shown in Table~\ref{tab:mws_sv1_areas}, this VAC includes not only observations that follow the regular DESI MWS strategy but also multiple dedicated fields placed on stellar streams, star clusters and dwarf galaxies. While a detailed exploration of the data from these observations is beyond the scope of this paper, here we provide a demonstration of measurements from fields centred on the Draco dwarf spheroidal galaxy.

Figure~\ref{fig:draco} shows the distribution of stars in the MWS VAC in [Fe/H] and radial velocity space in the vicinity of Draco. We only plot stars with Draco membership probability estimates from \citet{pace_li2022}, which were computed based on \gaia DR3 astrometry. The total number of stars in common with the membership catalogue is $\sim$ 950. The greyscale in the figure shows the 2D distribution of velocity and metallicity for these potential member stars. There is a broad distribution from halo contaminants and a peak associated with Draco stars near $V_\text{los} \sim -290\kms$ and $\feh\sim-2$. Red points in the figure show stars with a membership probability from \citet{pace_li2022} {$P_\mathrm{mem\_fixed\_complete}>0.5$}. Almost all high-probability member stars are clustered around Draco's radial velocity and metallicity, indicating that they are very likely Draco members. 

By looking at the metallicity distribution of Draco member stars in DESI, we find a median metallicity of $\feh \approx -2.25 \pm 0.03$, which at face value seems to differ significantly (by $\sim 0.3$\,dex) from the median metallicity of $-1.93\pm 0.01$ reported by \citet{kirby2011}. However, \citet{kirby2011} measured a significant metallicity gradient in Draco of $\approx -0.73 $\,dex\,deg$^{-1}$. The median angular distance of the member stars in the DESI sample from the centre of Draco is $\sim$ 0.25 degrees, as opposed to $\sim$ 0.1 degrees for the literature sample, owing to the large field of view of DESI. This metallicity gradient can, therefore, explain at least $\sim 0.1$\,dex of the discrepancy between \citet{kirby2011} and the DESI measurement. In fact, the median metallicity of the DESI members within 0.1 degrees is $-1.97$\,dex perfectly consistent with previous measurements.

The total number of Draco members detected in DESI is $\sim260$, which is comparable in size to the existing literature datasets on Draco \citep[see e.g.][]{kirby2010, walker2015}. However, our sample extends much further in radius, highlighting the power of the DESI survey for observations of the outskirts of dwarf galaxies.

\subsection{Wide binaries}

A final demonstration of the catalogue utilises an astrometrically selected set of wide binaries.

Since wide binaries are expected to share almost the same chemical compositions \citep{duchene2013} and have similar kinematics, these pairs should provide an additional internal consistency check on our stellar parameter measurements. To perform this test, we cross-match a subset of EDR VAC stars, selected to have good-quality measurements in radial velocities and metallicities, to the wide binary catalogue from \gaia EDR3 by \citet{El-Badry2021}. In addition to the selection criteria introduced in Section~\ref{sec:clean_sample_selection}, we impose the following requirements:
\begin{enumerate}
    \item {\tt CHISQ\_TOT} $< 5$
    \item {\tt SNR\_MED} $> 5$
    \item $\sigma(V_{\mathrm{rad}}) < 20 \,\mathrm{km\,\,s}^{-1}$
    \item $[\mathrm{Fe/H}]_{\mathrm{RVS}} > -3.9$ dex
    \item $[\mathrm{Fe/H}]_{\mathrm{SP}} > -4.9$ dex
    \item $R_{\mathrm{chance\_align}} < 0.1$.
\end{enumerate}
These criteria select $477$ high-confidence wide-binary pairs. We take the definition of ``primary" (P) and ``secondary" (S) from \citet{El-Badry2021}, which designate the brighter and fainter member in each pair, respectively, in the \gaia $G$ band. In a majority of pairs, both primary and secondary stars are on the main sequence and are K or M dwarfs. In the MWS VAC, all of these pairs have radial velocity and metallicity measurements from RVS, while metallicities measured from the SP pipeline are available for $453$ pairs.

\begin{figure*}
\includegraphics[width=\textwidth]{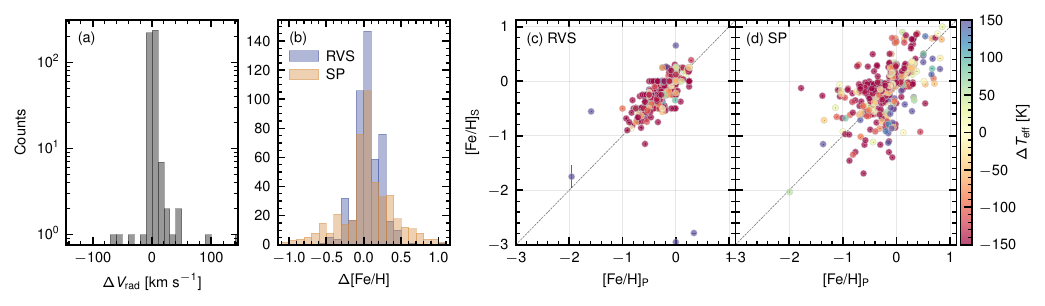}
\caption{Comparison of stellar parameters from the RVS and SP pipelines between the primary and secondary star for $477$ wide-binary pairs from \citet{El-Badry2021} with high confidence of being physical binaries ($R_{\mathrm{chance\_align}} < 0.1$). The panels show the distribution of differences in the radial velocities (a), the distribution of differences in metallicity measurements (b), and scatter plots (c and d) displaying metallicities colour-coded by temperature differences. Note that SP metallicities are available for $453$ pairs. 
\label{fig:wbs_rv}}
\end{figure*}

Figure~\ref{fig:wbs_rv} presents a comparison of stellar parameters from the RVS and SP pipelines. Panel (a) shows the distribution of $\Delta V_{\mathrm{rad}} = V_{\mathrm{rad,\,S}} - V_{\mathrm{rad,\,P}}$, the velocity difference between secondary and primary. Panel (b) presents the distribution of the metallicity difference, $\Delta [\mathrm{Fe/H}] =[\mathrm{Fe/H}]_{\mathrm{S}} - [\mathrm{Fe/H}]_{\mathrm{P}}$, while panels (c) and (d) are scatter plots of individual metallicity measurements, colour-coded by the differences in effective temperatures, i.e., $\Delta T_{\mathrm{eff}} = T_{\mathrm{eff, S}} - T_{\mathrm{eff, P}}$. 

The figure shows no significant difference between the radial velocity measurements of primary and secondary stars, with a small standard deviation of $ 1.84 \kms$ (measured from the 16th to 84th percentile range of the distribution). Some of that dispersion can be attributed to orbital motions and some to the individual radial velocity uncertainties, which are $\sim 2$ km\,s$^{-1}$ for this sample. 

Panels (b)-(d) in Figure~\ref{fig:wbs_rv} show that the distributions of $\Delta \mathrm{[Fe/H]}$ in both the RVS and SP pipelines have noticeable scatter between the measurements for primary and secondary stars.  Notably, however, the $\Delta \mathrm{[Fe/H]}$ values measured from the RVS pipeline have a standard deviation of $0.16$\,dex. This is comparable with the spread we obtained when comparing DESI measurements to APOGEE measurements. The distribution of $\Delta \mathrm{[Fe/H]}$ from the SP pipeline, however, shows a significantly wider spread with a dispersion of $0.31$\,dex. A similar picture can be seen in panels (c) and (d), where we plot the metallicity of the primary against the metallicity of the secondary for RVS and SP measurements. Here, larger scatter in SP metallicities is mostly caused by issues associated with cool stars, discussed in Section~\ref{sec:stellar_param_distribution} and the next section.

\section{Known limitations}
\label{sec:limitations}

In previous sections, we have described the contents of the MWS VAC, demonstrated the quality of DESI measurements and provided examples of scientific applications of the data. This section focuses on known issues and limitations in the catalogue that users should know. 

\begin{itemize}
    \item The distribution of parameters measured by {\tt RVSpecFit} shows concentrations at the edges and nodes of the underlying grid of stellar atmospheres. Since the code relies on multi-linear interpolation of the over-sampled PHOENIX grid, the \feh grid step of 0.25 dex leads to concentrations of measurements at  $\feh=-4,-3.75$ and so forth. Concentration at the grid points is more likely for bright stars. Despite this gridding, the overall accuracy of the [Fe/H] abundances is $\sim 0.1-0.2$ dex, as demonstrated in Section~\ref{sec:apogee_systematics}. The concentration of measurements around grid points is also observed for \logg, \teff and \alphafe. Because of this,  we believe that the uncertainties on stellar parameters in the RVS catalogue are likely underestimated and not well described by a Gaussian.
    \item Both RVS and SP metallicities display an excess of objects at the lower limit of their metallicity grids, $\feh=-4$ and $-5$, respectively. At least some of these measurements are caused by the spectra that cannot be fit by regular spectral models (i.e. they are non-stellar or otherwise unusual).
    \item The SP parameter distributions exhibit artificial concentrations at the boundary of the PHOENIX grid used for cool stars (see Figure \ref{fig:feh_afe_hist_rvs_sp}). 
    For this reason, we recommend avoiding using the SP parameters for stars analysed with that grid, as described in Appendix \ref{sec:sp_cuts}.
    \item The [$\alpha$/Fe] abundances for dwarfs cooler than about 4000~K  as measured by SP show concentration at values in the range $-0.1< \alphafe < 0.0$ and $-0.6 < \feh < -0.3$. We believe this is an artefact and discourage using the SP results for those stars, as described in Appendix \ref{sec:sp_cuts}.
    \item The carbon and magnesium abundances reported by the SP pipeline are unreliable. The reason carbon abundances are not trustworthy is because of an issue with the coating of the DESI blue collimators, which creates a throughput artefact in many spectra around 4300\AA\ \citep[see e.g. Figure 26 in ][]{abareshi2022}. This is the region of the CH absorption band, which is the most important indicator of carbon abundance in late-type stars at the resolution of DESI. 
    While the abundances of iron and calcium show a good correlation with the \feh and \alphafe parameters derived by the same pipeline, that is not the case for magnesium. 
    We, therefore, recommend not using the abundance of carbon or magnesium included in the EDR MWS VAC.
    \item When looking for high-velocity stars, the spectral classification by {\tt Redrock} cannot be used because {\tt Redrock} assumes stars have velocities $|V_\text{los}|<600 \kms$. That means that other criteria described in Section~\ref{sec:clean_sample_selection} should be used instead. However, these selections are more contaminated by non-stellar sources (QSOs and galaxies) than the criteria we recommend for general use.
    \item $v \sin i$ measurements of stellar rotation velocity provided by {\tt RVSpecFit} are problematic in this release. The rotational broadening measurement sometimes compensates for the inability of the model to fit some objects. For example, measurements of white dwarfs by the {\tt RVSpecFit} tend to have high inferred $v \sin i $. In addition, due to a software bug in the processing discovered during the preparation of this manuscript, $v \sin i $ measurements are biased to higher values for stars with higher line of sight velocities, $V_\text{los}>50\kms$ especially at low signal to noise. Thus, we recommend not to use $v \sin i$ values in this release. 
    
    \item Although the SP and RVS pipelines fit spectra of white dwarfs, these pipelines are not fully optimised for that purpose. SP uses both hydrogen-rich atmosphere (DA) and helium-rich atmosphere (DB) white dwarf model grids provided by \cite{Koester10}, whereas RVS does not have dedicated white dwarf models and instead uses regular stellar models and compensates for the pressure-broadened absorption features by increasing $v\sin i$ as noted above. While the majority of white dwarfs tend to have pure hydrogen DA-type spectra, many have more complex spectral features such as metal pollution, strong magnetic fields or molecular carbon from core dredge-up (see figure~2 of \citealt{Cooper23}). Dealing with these complications is beyond the scope of the SP and RVS pipelines. The DESI EDR white dwarf systems will be described separately in a future paper \citet{Manser2024}.

\end{itemize}

\section{Conclusions}

This paper presents the first value-added catalogue of stellar measurements based on the DESI Early Data Release, published in June 2023. The MWS VAC dataset includes over half a million spectra for more than 420,000 unique stars with a typical brightness $16 \lesssim r \lesssim 20$. We have described the different target types for the main DESI stellar survey and reviewed additional targets and special fields that were observed during survey validation and included in the EDR data set.

As described in \citet{Cooper23}, two stellar analysis pipelines have been developed for the DESI Milky Way Survey. These pipelines were run on the EDR dataset to obtain the radial velocities, chemical abundances and stellar parameters included in the MWS VAC.  By comparing these measurements with earlier high-quality reference catalogues, we conclude that our reported radial velocities are accurate to approximately $1 \kms$, metallicities to 0.15 dex, effective temperatures to 200 K, surface gravities to about 0.3 dex, and [$\alpha$/Fe] to about 0.1 dex. 

We have identified a number of limitations in this release of the MWS VAC, and provided a recommended set of criteria that can be used to select a high-quality subset of the data. At this point, the quality of the parameters measured by the RVS and SP pipelines is similar, each having its own benefits and drawbacks. We, therefore, do not provide a definite recommendation to use either set of parameters beyond the caveats listed in Section~\ref{sec:limitations}.

Despite some limitations in the value-added catalogue, it can already be used in many studies, including chemo-dynamical studies of the Milky Way halo and analyses of stellar streams and dwarf galaxies within the DESI EDR footprint.

As of the writing of this paper, DESI is progressing into its third year of observations, with regular data releases planned over the coming years. The stellar data included in the DESI EDR is a very small fraction of the total expected data set. Future data releases will include spectra and results for several million stars. Ongoing development of the analysis pipelines is expected to address the issues described in this paper, although few changes are expected in the data model and the target selection criteria. Therefore, we believe this value-added catalogue based on the EDR data set should be useful for the community, in particular to prepare for the scientific exploration of the full  DESI Milky Way Survey.  

\section*{Acknowledgments}

SK acknowledges support from the Science \& Technology Facilities Council (STFC) grant ST/Y001001/1. CAP acknowledges financial support from the Spanish Ministry MICINN projects AYA2017-86389-P and PID2020-117493GB-I00.
APC acknowledge support from a Taiwan Ministry of Education Yushan Fellowship and the Taiwan National Science and Technology Council grants 109-2112-M-007-011-MY3 and 112-2112-M-007-017-MY3.
T.S.L. acknowledges financial support from Natural Sciences and Engineering Research Council of Canada (NSERC) through grant RGPIN-2022-04794.
LBS and MV acknowledge support from NASA-ATP award 80NSSC20K0509. 

This material is based upon work supported by the U.S. Department of Energy (DOE), Office of Science, Office of High-Energy Physics, under Contract No. DE–AC02–05CH11231, and by the National Energy Research Scientific Computing Center, a DOE Office of Science User Facility under the same contract. Additional support for DESI was provided by the U.S. National Science Foundation (NSF), Division of Astronomical Sciences under Contract No. AST-0950945 to the NSF’s National Optical-Infrared Astronomy Research Laboratory; the Science and Technology Facilities Council of the United Kingdom; the Gordon and Betty Moore Foundation; the Heising-Simons Foundation; the French Alternative Energies and Atomic Energy Commission (CEA); the National Council of Science and Technology of Mexico (CONACYT); the Ministry of Science and Innovation of Spain (MICINN), and by the DESI Member Institutions: \url{https://www.desi.lbl.gov/collaborating-institutions}. Any opinions, findings, and conclusions or recommendations expressed in this material are those of the author(s) and do not necessarily reflect the views of the U. S. National Science Foundation, the U. S. Department of Energy, or any of the listed funding agencies.

The authors are honoured to be permitted to conduct scientific research on Iolkam Du’ag (Kitt Peak), a mountain with particular significance to the Tohono O’odham Nation.

This work used high-performance computing facilities operated by the Center for Informatics and Computation in Astronomy (CICA) at National Tsing Hua University. This equipment was funded by the Taiwan Ministry of Education, the Taiwan National Science and Technology Council, and National Tsing Hua University.  The authors thankfully acknowledge the technical expertise and assistance provided by the Spanish Supercomputing Network (Red Espa\~nola de Supercomputaci\'on), as well as the computing resources available at the Instituto de Astrof\'{\i}sica de Canarias, including the LaPalma Supercomputer and the compute servers of the Research Division.

For the purpose of open access, the author has applied a Creative Commons Attribution (CC BY) licence to any Author Accepted Manuscript version arising from this submission.

This paper made use of the Whole Sky Database (wsdb) created and maintained by Sergey Koposov at the Institute of Astronomy, Cambridge, with financial support from STFC and the European Research Council (ERC).

Software: {astropy\citep{2013A&A...558A..33A,2018AJ....156..123A,astropy2022}}


\vspace{5mm}

\section*{Data Availability}

All the data shown in the published graphs or necessary to reproduce them are available in a machine-readable form on Zenodo \url{https://doi.org/10.5281/zenodo.10498911}. The value-added catalogue itself is available at \url{https://data.desi.lbl.gov/doc/releases/edr/vac/mws/}.




\bibliographystyle{mnras}

\bibliography{references} 

\appendix 
\section{Selection criteria of the main MWS targets in SV3}
\change{
In this appendix we provide the selection criteria for the SV3 targets. The selection code is also available at \url{https://desitarget.readthedocs.io/en/latest/_modules/desitarget/sv3/sv3_cuts.html}.
We remark that the EDR covers DESI observations prior to the main survey. These targets were selected using {\it Gaia} DR2 photometry and astrometry, while the main survey selection uses {\it Gaia} DR3 astrometry. These selection criteria are slightly different from the main survey selection criteria described in \citet{Cooper23}.

All the columns used for selection of each target are available for each source in the VAC in the {\tt FIBERMAP} file extension (see Section~\ref{sec:dataproducts}). In the following selections $g$,$r$ refer to extinction corrected magnitudes obtained from Legacy Survey fluxes, $r_{obs}$ is the $r$ magnitude not corrected for extinction. Other variables are columns in the Legacy Survey table. $\pi,\sigma_\pi,\mu$ are \gaia parallax, parallax uncertainty and proper motion respectively.
All the main DESI classes, such as {\tt MWS\_MAIN\_BLUE}, {\tt MWS\_MAIN\_RED}, {\tt MWS\_BROAD} are selected using this common set of criteria: 
\begin{itemize}
\item $16 < r < 19$
\item $r_{obs} < 20$
\item type = PSF
\item {\tt gaia\_astrometric\_excess\_noise} < 3
\item {\tt gaia\_duplicated\_source }= False
\item {\tt brick\_primary} = True
\item {\tt nobs\_{g,r}} > 0
\item {g,r}\_flux > 0
\item {\tt fracmasked\_{g,r}} < 0.5
\end{itemize}

Definition of {\tt MWS\_MAIN\_BLUE}:
\begin{itemize}
    \item $g-r<0.7$
    \item {\tt astrometric\_params\_solved}=31
\end{itemize}

Definition of {\tt MWS\_MAIN\_RED}:

\begin{itemize}
    \item $g-r>=0.7$
    \item {\tt astrometric\_params\_solved} =31
    \item $\pi < max(3 \sigma_\pi , 1) $
    \item $|\mu| <  7$
\end{itemize}

Definition of {\tt MWS\_BROAD}:
\begin{itemize}
    \item $g-r>=0.7$
    \item {\tt astrometric\_params\_solved}$< $ 31 or $\pi > max(3 \sigma_\pi , 1) $ or  $|\mu| > 7 $ 
\end{itemize}

}
\begin{table*}
\begin{tabular}{ccc}
\hline
Name & Units & Description \\
\hline
VRAD & \kms & Radial velocity \\
VRAD\_ERR & \kms & Radial velocity error \\
VRAD\_SKEW &  & Radial velocity posterior skewness \\
VRAD\_KURT &  & Radial velocity posterior kurtosis \\
LOGG &  & Logarithm of surface gravity \\
TEFF & K & Effective temperature \\
ALPHAFE &  & \alphafe from template fitting \\
FEH &  & \feh from template fitting \\
LOGG\_ERR &  & Log of surface gravity uncertainty \\
TEFF\_ERR & K & Effective temperature uncertainty \\
ALPHAFE\_ERR &  & \alphafe uncertainty from template fitting \\
FEH\_ERR &  & \feh uncertainty from template fitting \\
VSINI & \kms & Stellar rotation velocity \\
NEXP &  &  Number of exposures (unused) \\
CHISQ\_TOT &  & Total chi-square for all arms \\
CHISQ\_C\_TOT &  & Total chi-square for all arms for polynomial only fit \\
CHISQ\_B &  & Chi-square in the B arm \\
CHISQ\_C\_B &  & Chi-square in the B arm after fitting continuum only \\
CHISQ\_R &  & Chi-square in the R arm \\
CHISQ\_C\_R &  & Chi-square in the R arm after fitting continuum only \\
CHISQ\_Z &  & Chi-square in the Z arm \\
CHISQ\_C\_Z &  & Chi-square in the Z arm after fitting continuum only \\
RVS\_WARN &  & {\tt RVSpecFit} warning flag \\
REF\_ID &  & Tyc1*1,000,000+Tyc2*10+Tyc3 for Tycho-2; source\_id for \gaia DR2 \\
REF\_CAT &  & Reference catalog source for star: T2 for Tycho-2
\\ & &  G2 for \gaia DR2, L2 for the SGA, empty otherwise \\
TARGET\_RA & deg & Target right ascension \\
TARGET\_DEC & deg & Target declination \\
TARGETID &  & Unique DESI target ID \\
SN\_B &  & Median S/N in the B arm \\
SN\_R &  & Median S/N in the R arm \\
SN\_Z &  & Median S/N in the Z arm \\
SUCCESS &  & Did {\tt RVSpecFit} succeed ({\tt RVS\_WARN}=0) or fail \\
\hline
\end{tabular}
\caption{The list of columns in the {\tt RVTAB} extension with measurements from {\tt RVSpecFit}.}
\label{tab:rvtab}
\end{table*}

\begin{table*}
\begin{tabular}{ccc}
\hline
Name & Units & Description \\
\hline
SUCCESS &  & Boolean flag indicating whether the code has likely produced useful results \\
TARGETID &  & Unique DESI target ID \\
TARGET\_RA & deg & Target right ascension \\
TARGET\_DEC & deg & Target declination \\
REF\_ID &  & Tyc1*1,000,000+Tyc2*10+Tyc3 for Tycho-2; source ID for \gaia DR2 \\
REF\_CAT &  & Reference catalog source for star: T2 for Tycho-2
\\ & &  G2 for \gaia DR2, L2 for the SGA, empty otherwise \\
SRCFILE &  & DESI data file \\
BESTGRID &  & Model grid that produced the best fit \\
TEFF & K & Effective temperature \\
LOGG &  & Logarithm of surface gravity in $cm/s^2$ \\
FEH &  & Metallicity \feh
\\
ALPHAFE &  & $\alpha$-to-iron ratio \alphafe \\
LOG10MICRO &  & Log10 of Microturbulence (km/s) \\
PARAM &  & Array of atmospheric parameters (\feh, \alphafe, log10micro, \teff, \logg) \\
COVAR &  & Covariance matrix for (\feh, \alphafe, log10micro, \teff, \logg) \\
ELEM &  & Elemental abundance ratios to hydrogen [elem/H] \\
ELEM\_ERR &  & Uncertainties in the elemental abundance ratios \\
CHISQ\_TOT &  & Total $\chi^2$ \\
SNR\_MED &  & Median signal-to-ratio \\
RV\_ADOP & \kms & Adopted Radial Velocity \\
RV\_ERR & \kms & Uncertainty in the adopted Radial Velocity \\
HEALPIX &  & HEALPix pixel containing this location at NSIDE=64 in the NESTED scheme \\
\hline
\end{tabular}
\caption{The list of columns in the {\tt SPTAB} extension with measurements from {\tt FERRE}.}
\label{tab:sptab}
\end{table*}

\appendix 
\section{Recommended selection for the SP parameters}
\label{sec:sp_cuts}

In the case of the results of the SP pipeline, included in the {\tt SPTAB} extensions of various files, we recommend the following cuts to obtain a science-grade sample: 
\begin{itemize}
    \item ${\tt SNR\_MED} > 5$, a minimum median signal-to-noise ratio of 5, 
    \item ${\tt CHISQR\_TOT} < 5$, a reduced $\chi^2$ smaller than 5, 
    \item ${\tt TEFF} > 4000$ K, to avoid M-type stars, 
    \item ${\tt FEH} > -4.9$, avoiding the stars that accumulate the edge of the grids, and 
    \item ${\tt BESTGRID} \neq$ {\tt s\_rdesi1}, avoiding the stars that were fitted with the PHOENIX grid.
\end{itemize}

The use of the PHOENIX grid for stars with effective temperatures between 2300 K and 5000 K was adopted to supplement the \citet{kurucz2005} grids since the latter only cover $T_{\rm eff} \ge 3500$ K. For stars in the overlapping range of temperatures between two grids the results with the lowest $\chi^2$ were adopted. We find that this led to spurious results when the PHOENIX grid was used. One such example is the accumulation of stars near the grid edge at 5000\,K illustrated in Figure~\ref{fig:params}, the presence of systematic errors in the surface gravity of cool dwarfs, or large scatter in the abundances for cool dwarfs seen in Figure~\ref{fig:wbs_rv}.

\section{Recommended selection for the RVS parameters}
\label{sec:rvs_cuts}

On top of the selection listed in Section~\ref{sec:clean_sample_selection}, for a more reliable subset of RVS parameters, we recommend applying the following cuts. 
\begin{itemize}
\item  {\tt FEH}>-3.9 removes objects clumped at the bottom edge of the metallicity grid.
\end{itemize}

 
\section{Author affiliations}
\label{sec:affiliations}
$^{1}$ Institute for Astronomy, University of Edinburgh, Royal Observatory, Blackford Hill, Edinburgh EH9 3HJ, UK\\
$^{2}$ Institute of Astronomy, University of Cambridge, Madingley Road, Cambridge CB3 0HA, UK\\
$^{3}$ Kavli Institute for Cosmology, University of Cambridge, Madingley Road, Cambridge CB3 0HA, UK\\
$^{4}$ Instituto de Astrof\'{i}sica de Canarias, C/ Vía L\'{a}ctea, s/n, E-38205 La Laguna, Tenerife, Spain \\  Universidad de La Laguna, Dept. de Astrof\'{\i}sica, E-38206 La Laguna, Tenerife, Spain\\
$^{5}$ Institute of Astronomy and Department of Physics, National Tsing Hua University, 101 Kuang-Fu Rd. Sec. 2, Hsinchu 30013, Taiwan\\
$^{6}$ Department of Astronomy \& Astrophysics, University of Toronto, Toronto, ON M5S 3H4, Canada\\
$^{7}$ Department of Astronomy, University of Michigan, Ann Arbor, MI 48109, USA\\
$^{8}$ Steward Observatory, University of Arizona, 933 N, Cherry Ave, Tucson, AZ 85721, USA\\
$^{9}$ Institute for Computational Cosmology, Department of Physics, Durham University, South Road, Durham DH1 3LE, UK\\
$^{10}$ NSF's NOIRLab, 950 N. Cherry Ave., Tucson, AZ 85719, USA\\
$^{11}$ Astrophysics Group, Department of Physics, Imperial College London, Prince Consort Rd, London, SW7 2AZ, UK\\
$^{12}$ Department of Physics, University of Warwick, Gibbet Hill Road, Coventry, CV4 7AL, UK\\
$^{13}$ Physics Department, Yale University, P.O. Box 208120, New Haven, CT 06511, USA\\
$^{14}$ Department of Astronomy and Astrophysics, UCO/Lick Observatory, University of California, 1156 High Street, Santa Cruz, CA 95064, USA\\
$^{15}$ Department of Astronomy and Astrophysics, University of California, Santa Cruz, 1156 High Street, Santa Cruz, CA 95065, USA\\
$^{16}$ University of California Observatories, 1156 High Street, Sana Cruz, CA 95065, USA\\
$^{17}$ University of Michigan, Ann Arbor, MI 48109, USA\\
$^{18}$ Lawrence Berkeley National Laboratory, 1 Cyclotron Road, Berkeley, CA 94720, USA\\
$^{19}$ Physics Dept., Boston University, 590 Commonwealth Avenue, Boston, MA 02215, USA\\
$^{20}$ Department of Physics \& Astronomy, University College London, Gower Street, London, WC1E 6BT, UK\\
$^{21}$ Instituto de F\'{\i}sica, Universidad Nacional Aut\'{o}noma de M\'{e}xico,  Cd. de M\'{e}xico  C.P. 04510,  M\'{e}xico\\
$^{22}$ Department of Physics \& Astronomy and Pittsburgh Particle Physics, Astrophysics, and Cosmology Center (PITT PACC), University of Pittsburgh, 3941 O'Hara Street, Pittsburgh, PA 15260, USA\\
$^{23}$ Departamento de F\'isica, Universidad de los Andes, Cra. 1 No. 18A-10, Edificio Ip, CP 111711, Bogot\'a, Colombia\\
$^{24}$ Observatorio Astron\'omico, Universidad de los Andes, Cra. 1 No. 18A-10, Edificio H, CP 111711 Bogot\'a, Colombia\\
$^{25}$ Institut d'Estudis Espacials de Catalunya (IEEC), 08034 Barcelona, Spain\\
$^{26}$ Institute of Cosmology \& Gravitation, University of Portsmouth, Dennis Sciama Building, Portsmouth, PO1 3FX, UK\\
$^{27}$ Institute of Space Sciences, ICE-CSIC, Campus UAB, Carrer de Can Magrans s/n, 08913 Bellaterra, Barcelona, Spain\\
$^{28}$ Sorbonne Universit\'{e}, CNRS/IN2P3, Laboratoire de Physique Nucl\'{e}aire et de Hautes Energies (LPNHE), FR-75005 Paris, France\\
$^{29}$ Departament de F\'{i}sica, Serra H\'{u}nter, Universitat Aut\`{o}noma de Barcelona, 08193 Bellaterra (Barcelona), Spain\\
$^{30}$ Institut de F\'{i}sica d’Altes Energies (IFAE), The Barcelona Institute of Science and Technology, Campus UAB, 08193 Bellaterra Barcelona, Spain\\
$^{31}$ Instituci\'{o} Catalana de Recerca i Estudis Avan\c{c}ats, Passeig de Llu\'{\i}s Companys, 23, 08010 Barcelona, Spain\\
$^{32}$ Department of Physics and Astronomy, Siena College, 515 Loudon Road, Loudonville, NY 12211, USA\\
$^{33}$ National Astronomical Observatories, Chinese Academy of Sciences, A20 Datun Rd., Chaoyang District, Beijing, 100012, P.R. China\\
$^{34}$ IRFU, CEA, Universit\'{e} Paris-Saclay, F-91191 Gif-sur-Yvette, France\\
$^{35}$ Department of Physics and Astronomy, University of Waterloo, 200 University Ave W, Waterloo, ON N2L 3G1, Canada\\
$^{36}$ Perimeter Institute for Theoretical Physics, 31 Caroline St. North, Waterloo, ON N2L 2Y5, Canada\\
$^{37}$ Waterloo Centre for Astrophysics, University of Waterloo, 200 University Ave W, Waterloo, ON N2L 3G1, Canada\\
$^{38}$ Department of Physics, Kansas State University, 116 Cardwell Hall, Manhattan, KS 66506, USA\\
$^{39}$ Department of Physics and Astronomy, Sejong University, Seoul, 143-747, Korea\\
$^{40}$ CIEMAT, Avenida Complutense 40, E-28040 Madrid, Spain\\
$^{41}$ Space Telescope Science Institute, 3700 San Martin Drive, Baltimore, MD 21218, USA\\
$^{42}$ Department of Physics, University of Michigan, Ann Arbor, MI 48109, USA\\
\bsp	
\label{lastpage}
\end{document}